\documentclass[showpacs,preprintnumbers,amsmath,amssymb]{revtex4}
\usepackage{amsmath,amssymb,graphics,epsfig,subfigure}
\usepackage{color}

\begin{document}
\renewcommand{\baselinestretch}{1.3}

\title{Testing the microstructure of $d$-dimensional charged Gauss-Bonnet anti-de Sitter black holes}

\author{Shao-Wen Wei \footnote{weishw@lzu.edu.cn, corresponding author},
  Yu-Xiao Liu \footnote{liuyx@lzu.edu.cn}}

\affiliation{Lanzhou Center for Theoretical Physics, Key Laboratory of Theoretical Physics of Gansu Province, School of Physical Science and Technology, Lanzhou University, Lanzhou 730000, People's Republic of China,\\
 Institute of Theoretical Physics $\&$ Research Center of Gravitation,
Lanzhou University, Lanzhou 730000, People's Republic of China,\\
 Joint Research Center for Physics, Lanzhou University and Qinghai Normal University, Lanzhou 730000 and Xining 810000, China}

\begin{abstract}
Understanding black hole microstructure via the thermodynamic geometry can provide us with more deeper insight into black hole thermodynamics in modified gravities. In this paper, we study the black hole phase transition and Ruppeiner geometry for the $d$-dimensional charged Gauss-Bonnet anti-de Sitter black holes. The results show that the small-large black hole phase transition is universal in this gravity. By reducing the thermodynamic quantities with the black hole charge, we clearly exhibit the phase diagrams in different parameter spaces. Of particular interest is that the radius of the black hole horizon can act as the order parameter to characterize the black hole phase transition. We also disclose that different from the five-dimensional neutral black holes, the charged ones allow the repulsive interaction among its microstructure for small black hole of higher temperature. Another significant difference between them is that the microscopic interaction changes during the small-large black hole phase transition for the charged case, where the black hole microstructure undergoes a sudden change. These results are helpful for peeking into the microstructure of charged black holes in the Gauss-Bonnet gravity.
\end{abstract}



\maketitle

\section{Introduction}

Extended thermodynamics is one of the most active areas in black hole physics. In particular, in the extended phase space, the cosmological constant is interpreted as the pressure and its conjugate quantity is treated as the thermodynamic volume \cite{Kastor}. Although the black hole phase transition has been observed in Refs. \cite{Chamblin,Chamblin2,Cognola}. The precise analogy between the black hole and the van der Waals (VdW) fluid was completed in \cite{Kubiznak}. Reminiscent of the liquid-gas phase transition, the small-large black hole phase transition was found to universally exist in the anti-de Sitter (AdS) black holes. Besides, more interesting phase transition, such as the reentrant phase transition, triple point, superfluid phase were observed revealing that black hole has rich thermodynamic phenomena \cite{Kubiznak,Altamirano,AltamiranoKubiznak,Altamirano3,Dolan,Liu,Wei2,Frassino,
Cai,XuZhao,Hennigar,Hennigar2,Tjoa2,Ruihong}.

As we know, black holes have well defined macroscopic thermodynamic quantities \cite{Hawking,Bekensteina,Bekensteinb}, such as the temperature, entropy and so on. Four thermodynamic laws of black holes also hold \cite{Bardeen}. However, black hole microstructure is a huge challenge. String theory and fuzzball model showed some fruitful results, while plenty of fundamental problems are still waiting for solved. How to test the microscopic properties of black hole by using these macroscopic thermodynamic quantities is unknown.

Ruppeiner geometry developed in \cite{Ruppeiner} provides us with a natural approach to peek into the microstructure of a thermodynamic system. It starts with the Boltzmann entropy formula, and constructs the corresponding geometry in a parameter space. Applying the similar treatment with the Riemannian geometry, the scalar curvature can be produced. After the systematic study of a number of fluid systems, some observations underlying the geometry were uncovered. Here we summarize them as follows \cite{Ruppeiner}: i) The line element of the geometry measures the distance between two neighbouring fluctuation states, and the less probable a fluctuation between two neighboring states, the longer the distance between them;  ii) The sign of the scalar curvature can be used to test the interaction among the micromolecules of a thermodynamic system. Positive and negative scalar curvatures are related to repulsive and attractive interactions, respectively; iii) The scalar curvature of the geometry corresponds to the correlation length near a critical point; iv) Critical phenomena exists for such geometry.

Ruppeiner geometry has been introduced in the study of the black hole physics \cite{CaiCho}. However, as we shown above, it is closely related with the study of the phase transition, in particular the VdW-like one. However, most early studies focus on the divergent behaviors of the scalar curvature, as well as that of the heat capacity.

In the extended phase space, the small-large black hole phase transition was observed in many black hole backgrounds. At the critical point, the black holes and the VdW fluid share the same critical exponents. It is extensively known that phase transition is resulted by the coexistence and competition of the microcosmic constituents of a thermodynamic system. So the microscopic properties of the black hole can be exposed via the phase transition. More importantly, this provides an excellent opportunity for the Ruppeiner geometry.

Combining with the small-large black hole phase transition in a charged AdS black hole background, the Ruppeiner geometry was constructed. Via the corresponding scalar curvature, the black hole microstructure was first tested \cite{Weiw}. Further, we reconstructed the Ruppeiner geometry for the black hole systems in a more natural approach \cite{Weiwa2,WeiWeiWei}. First, we applied it to the VdW fluid. Excluding the coexistence region where the equation of state does not hold, we found that there is only the attractive interaction among the fluid molecules. However, when applying it to the four-dimensional charged AdS black hole, the vanishing heat capacity will spoil the geometry. To solve this problem, we introduced a new normalized scalar curvature. It suggests that repulsive interaction can exist for the small black hole with high temperature, quite different from the VdW fluid. This uncovers the particular microstructure of the black hole. Moreover, the critical phenomena of the normalized scalar curvature were observed. Generalizing the study to higher dimensions, we obtained the similar results. Subsequently, the approach has been extended to other black hole backgrounds, and some interesting microscopic properties of the black hole microstructure were disclosed \cite{Dehyadegari,Zangeneh:2016fhy,Moumni,Miao,Du,Xuz,GhoshBhamidipati,Kumara2020,Yerra:2020oph,Wu:2020fij,
Vaid,Rizwan,Mansoori,Kumara,Mannw,Dehyadegariw,Wei2020d,Wei2020a,Zhourun,YPhu,Xuzzm}.

On the other hand, the Gauss-Bonnet (GB) gravity is a widely concerned modified
gravity. The thermodynamics and phase transition of the black holes were extensively investigated \cite{Liu,Wei2,Cai}. In Refs. \cite{Liu,Wei2}, we first observed the triple point in six-dimensional charged GB-AdS black holes. While in other dimensions, only the small-large black hole phase transition was found. More interestingly, for the five-dimensional neutral GB-AdS black hole, the coexistence curve of the small and large black holes was analytically obtained \cite{Mol}. For the charged case, the analytic coexistence curve is gained in the grand canonical ensemble \cite{Zhourun}. These provide us with a good opportunity to exactly study its thermodynamics.

For the black holes in the four-dimensional GB gravity, we found that there are the small-large black hole phase transition both in the canonical and grand canonical ensembles \cite{Wei2020d}. After constructing the Ruppeiner geometry, we found that its microscopic properties are similar to the charged AdS black hole in Einstein gravity. For the small black hole with high temperature, the repulsive interaction dominates among the microstructure.

However, for the five-dimensional neutral GB-AdS black hole, the result behaves quite interesting \cite{Wei2020a}. When excluding the coexistence regions, only the attractive interaction is allowed. This suggests that the repulsive interaction is not necessary for the black holes. Making use the analytical coexistence curve, we observed that along the coexistence small and large black hole curves, the normalized scalar curvature keeps the same values. As we know, when a phase transition takes place, the microstructure gets a sudden change. The value of the scalar curvature is generally believed to indicate the interaction, and same values reflect the same interaction. So for the five-dimensional neutral black hole, a significant property reflected from the Ruppeiner geometry is that its microscopic interaction keeps the same while its microstructure suddenly changes. This uncovers the unique nature of the GB gravity. This particular property was also observed in the grand canonical ensemble \cite{Zhourun}.

Since the charged case in the GB gravity has not been studied in $d\geq$5 in the canonical ensemble, in this paper, we aim to study the thermodynamics and microstructure for them and to see whether an interesting different microstructure can be revealed. The present paper is organized as follows. In Sec. \ref{iner}, we briefly review the thermodynamics and Ruppeiner geometry for the charged GB-AdS black holes. In Sec. \ref{ptpsifd}, we consider the black hole in five dimensions. The phase diagram and the scalar curvature are given. The critical phenomena are also numerically investigated. Then in Sec. \ref{ptinfd}, we generalize the study to the black holes in $d\geq$6. Finally, we summarize and discuss our results.

\section{Thermodynamics and microstructure of charged Gauss-Bonnet AdS black holes}
\label{iner}

In this section, we shall introduce the thermodynamics and Ruppeiner geometry for the $d$-dimensional charged GB-AdS black holes.

\subsection{Thermodynamics}

The line element for the black hole is \cite{Boulware,Cai2,Wiltshire,Cvetic2}
\begin{eqnarray}
 ds^2&=&-f(r)dt^2+\frac{1}{f(r)}dr^2+r^2d\Omega^2_{d-2},\\
 f(r)&=&1+\frac{r^2}{2\alpha}\left(1-\sqrt{1+\frac{64\pi\alpha M}{(d-2)\omega_{d-2} r^{d-1}}-\frac{8\alpha Q^2}{(d-2)(d-3)r^{2d-4}}-\frac{64\pi\alpha P}{(d-1)(d-2)}}\right),
\end{eqnarray}
where $M$ and $Q$ are the mass and charge of the black hole, respectively. The parameter $\alpha$ is related with the GB coupling $\alpha=(d-3)(d-4)\alpha_{\text{GB}}$ with dimension [length]$^2$. The volume of the unit $(d-2)$-sphere is $\omega_{d-2}=2\pi^{(d-1)/2}/\Gamma((d-1)/2)$. Here we consider the thermodynamic quantity in the extended phase space, where the cosmological constant $\Lambda$ has been interpreted as the pressure $P=-\frac{\Lambda}{8\pi}$. This solution is obtained from the following action
\begin{eqnarray}
 S=\frac{1}{16\pi}\int d^{d}x\sqrt{-g}\left(R-2\Lambda+\alpha_{\text{GB}}\mathcal{L}_{\text{GB}} -4\pi F_{\mu\nu}F^{\mu\nu}\right),
\end{eqnarray}
where the GB Lagrangian $\mathcal{L}_{\text{GB}}=\mathcal{R}_{\mu\nu\gamma\delta}\mathcal{R}^{\mu\nu\gamma\delta}
-4\mathcal{R}_{\mu\nu}\mathcal{R}^{\mu\nu}+\mathcal{R}^{2}$.

The radius of the event horizon can be obtained by solving $f(r_{\text{h}})=0$, which depends on the black hole mass, charge and pressure. Here we can also express the black hole mass as
\begin{eqnarray}
 M=\frac{(d-2)\omega_{d-2}r_{\text{h}}^{d-3}}{16\pi}\left(1+\frac{\alpha}{r_{\text{h}}^{2}}
  +\frac{16\pi P r_{\text{h}}^2}{(d-1)(d-2)}\right)+\frac{\omega_{d-2}Q^{2}}{8\pi(d-3)r_{\text{h}}^{d-3}}.
\end{eqnarray}
Other thermodynamic quantities, such as the temperature $T$, entropy $S$, electric potential $\Phi$, and conjugate quantity $\mathcal{A}$ to $\alpha$ are given by \cite{Cai}
\begin{eqnarray}
 T&=&\frac{-2Q^2 r_{\text{h}}^{7-2 d}+2 (d-3) (d-2) r_{\text{h}}+32 \pi  P
   r_{\text{h}}^3}{4 \pi  (d-2) \left(2 \alpha
   +r_{\text{h}}^2\right)}-\frac{5 \alpha -\alpha  d}{8 \pi
   \alpha  r_{\text{h}}+4 \pi  r_{\text{h}}^3},\label{tem}\\
 S&=&\frac{\omega_{d-2}r_{\text{h}}^{d-2}}{4}\left(1+\frac{2(d-2)\alpha}{(d-4)r_{\text{h}}^2}\right),\\
 \Phi&=&\frac{\omega_{d-2}Q}{4\pi(d-3)r_{\text{h}}^{d-3}},\\
 \mathcal{A}&=&-\frac{\omega_{d-2}  r_{\text{h}}^{-d-5} \left(r_{\text{h}}^{2 d} \left(-2
   \alpha  (d-2)+(d-2)^2 r_{\text{h}}^2+32 \pi  P
   r_{\text{h}}^4\right)-4Q^2 r_{\text{h}}^8\right)}{16 \pi  (d-4)
   \left(2 \alpha +r_{\text{h}}^2\right)}.
\end{eqnarray}
The specific volume and  thermodynamic volume are given by
\begin{eqnarray}
 v&=&\frac{4r_{\text{h}}}{d-2},\\
 V&=&\frac{\omega_{d-2}r_{\text{h}}^{d-1}}{d-1}.
\end{eqnarray}
It is clear that $V\sim v^{d-1}$, so they are not linear, which is quite different from the VdW fluid. For the black hole thermodynamics, these two volumes have different applications, see the related discussion in Ref. \cite{Wei3}. One can easily confirm that the first law and the Smarr formula hold
\begin{eqnarray}
 dM&=&TdS+\Phi dQ+\mathcal{A}d\alpha+VdP,\label{Firstlaw}\\
 (d-3)M&=&(d-2)TS+(d-3)Q\Phi +2\mathcal{A}\alpha-2PV.
\end{eqnarray}
From (\ref{Firstlaw}), we find that the pressure-volume term is $VdP$ rather than $-PdV$. So, as suggested in Ref. \cite{Kastor}, the black hole mass $M$ here should be treated as the enthalpy $H$ of the system.

Solving Eq. (\ref{tem}), we can obtain the equation of state
\begin{eqnarray}
 P=\frac{T}{v}-\frac{d-3}{(d-2)\pi v^{2}}+\frac{32T\alpha}{(d-2)^2v^3}
 -\frac{16\alpha(d-5)}{(d-2)^3\pi v^4}
 +\frac{2^{4d-11}Q^2}{\pi (d-2)^{2d-4}v^{2d-4}}.
\end{eqnarray}
Obviously, this equation closely depends on the GB coupling $\alpha$ and charge $Q$. According to the classification of the black hole quantities \cite{Wei6}, this black hole system is a double characteristic parameter system. However, we can reduce these thermodynamic quantities with the charge $Q$. This shall help us to simplify the study. Via the dimensional analysis, we adopt the following reduced forms
\begin{eqnarray}
 \tilde{T}=TQ^{\frac{1}{d-3}},\quad
 \tilde{P}=PQ^{\frac{2}{d-3}},\quad
 \tilde{r}_{\text{h}}=r_{\text{h}}Q^{-\frac{1}{d-3}},\quad
 \tilde{v}=vQ^{-\frac{1}{d-3}},\quad
 \tilde{V}=VQ^{-\frac{d-1}{d-3}}.
\end{eqnarray}
Then the reduced equation of state reads
\begin{eqnarray}
 \tilde{P}=\frac{\tilde{T}}{\tilde{v}}-\frac{d-3}{(d-2)\pi \tilde{v}^{2}}+\frac{32\tilde{T}\tilde{\alpha}}{(d-2)^2\tilde{v}^3}
 -\frac{16\tilde{\alpha}(d-5)}{(d-2)^3\pi \tilde{v}^4}
 +\frac{2^{4d-11}}{\pi (d-2)^{2d-4}\tilde{v}^{2d-4}}.
\end{eqnarray}
Now we find that the equation of state only depends on the reduced GB coupling $\tilde{\alpha}=\alpha/Q^{2/(d-3)}$. Moreover, we can express the reduced pressure in terms of the thermodynamic volume $\tilde{V}$. Another important thermodynamic quantity is the Gibbs free energy, which is
\begin{eqnarray}
 \tilde{G}=\frac{H-TS}{Q}&=&\frac{\omega_{d-2}}{16\pi(d-4)(d-3)(d-2)
   (d-1)\left(2 \alpha +\tilde{r}_{\text{h}}^2\right)\tilde{r}_{\text{h}}^{d+5}}\times
   \bigg((d-3) \tilde{r}_{\text{h}}^{2d}\bigg(2\tilde{\alpha}^2(d-2)^2
   (d-1)\nonumber\\
   &+&\tilde{\alpha}(d-2)\tilde{r}_{\text{h}}^2\left((d-9)d-96\pi \tilde{P}
   \tilde{r}_{\text{h}}^2+8\right)+(d-4)\tilde{r}_{\text{h}}^4 \left((d-3) d-16 \pi
   \tilde{P} \tilde{r}_{\text{h}}^2+2\right)\bigg)\nonumber\\
   &+&2(d-1) \tilde{r}_{\text{h}}^8 \left(2\tilde{\alpha}(d-2)(2d-7)+(d-4)(2d-5)\tilde{r}_{\text{h}}^2\right)\bigg).
\end{eqnarray}
The local thermodynamic stability is measured by the heat capacity. Positive or negative one indicates the system is local stable or unstable. The heat capacity can be calculated by
\begin{eqnarray}
 C_{Q}=T\left(\frac{\partial S}{\partial T}\right)_{Q, \alpha, P}.
\end{eqnarray}
Since it is in a complex form, we shall not show it here. In the following section, we will investigate the phase structure for different dimension number $d$.

\subsection{Ruppeiner geometry and microstructure}

In this subsection, we would like to give a brief introduction for the Ruppeiner geometry, and to show how black hole microstructure is uncovered via the geometry.

Here we consider that the system S we concerned is surrounded by its huge environment E. Then the total entropy of the complete system can be written as
\begin{equation}
 S(x^{0},x^{1})=S_{\rm S}(x^{0},x^{1})+S_{\rm E}(x^{0},x^{1}),
\end{equation}
where we assume the system has two independent thermodynamic extensive variables $x^0$ and $x^1$. Moreover, $S_{\rm S}\ll S_{\rm E}\sim S$ is required. Near the local maximum of the entropy at $x^{\mu} = x^{\mu}_0$, we expand the total entropy as
\begin{eqnarray}
 S&=&S_{0} + \left. \frac{\partial S_{\rm S}}{\partial x^{\mu}}  \right|_{x^{\mu}_0} \Delta x^{\mu}_{\rm S}
      + \left.\frac{\partial S_{\rm E}}{\partial x^{\mu}}   \right|_{x^{\mu}_0}   \Delta x^{\mu}_{\rm E}\nonumber\\
 &&   + \left. \frac{1}{2}\frac{\partial^{2}S_{\rm S}}{\partial x^{\mu}\partial x^{\nu}}
        \right|_{x^{\mu}_0}  \Delta x^{\mu}_{\rm S}\Delta x^{\nu}_{\rm S}
      + \left. \frac{1}{2}\frac{\partial^{2}S_{\rm E}}{\partial x^{\mu}\partial x^{\nu}}
        \right|_{x^{\mu}_0}  \Delta x^{\mu}_{\rm E}\Delta x^{\nu}_{\rm E}
   +\cdots, \quad (\mu, \nu=0,1),
 \end{eqnarray}
where we denote the local maximum of the entropy $S(x^{\mu}_0)$ with $S_{0}$. Since the fluctuating parameters are additive, i.e., $x_{S}^{\mu}+x_{E}^{\mu}=x_{total}^{\mu}=constant$, one can easily get $\left.\frac{\partial S_{\rm S}}{\partial x^{\mu}} \right|_{0} \Delta x^{\mu}_{\rm S}
   + \left.\frac{\partial S_{\rm E}}{\partial x^{\mu}} \right|_{0} \Delta x^{\mu}_{\rm E}=0$. Thus we have $\Delta S=S-S_{0}$,
\begin{equation}
 \Delta S=\left. \frac{1}{2}\frac{\partial^{2}S_{\rm S}}{\partial x^{\mu} \partial x^{\nu}}
              \right|_{0} \Delta x^{\mu}_{\rm S} \Delta x^{\nu}_{\rm S}
          +   \left. \frac{1}{2}\frac{\partial^{2}S_{\rm E}}{\partial x^{\mu} \partial x^{\nu}}
              \right|_{0}  \Delta x^{\mu}_{\rm E} \Delta x^{\nu}_{\rm E}
          +   \cdots\;.
 \end{equation}
The second term is much smaller than the first term, so we ignore it. Finally, the probability of finding the system in the internals $x_0\sim (x_0 + \Delta x_0)$ and $x_1\sim (x_1 + \Delta x_1)$ can be expressed in the following form
\begin{eqnarray}
 P(x^{0},x^{1}) \propto e^{-\frac{1}{2}\Delta l^{2}},
 \end{eqnarray}
 where
 \begin{eqnarray}
  \Delta l^{2}&=&-\frac{1}{k_{\rm B}}g_{\mu\nu} \Delta x^{\mu}\Delta x^{\nu},\label{Ds}\\
 g_{\mu\nu}&=& \left.\frac{\partial^{2}S_{\rm S}}{\partial x^{\mu}\partial x^{\nu}} \right|_{0} .\label{gmunu}
 \end{eqnarray}
According to the thermodynamic information geometry, $\Delta l^{2}$ measures the distance between two neighboring fluctuation states. The less probable a fluctuation between two neighboring states, the longer the distance between them. Following the Riemannian geometry, we can construct the corresponding scalar curvature for this Ruppeiner geometry. The empirical observation suggests that positive (negative) scalar curvature indicates the repulsive (attractive) interaction among the small molecules of the system. Further study also uncovers that the scalar curvature is related to the correlation length for the thermodynamic system near the critical point.

Taking the internal energy $U$ and thermodynamic volume $V$ as the fluctuation coordinates, the line element can be reexpressed via the first law of the black hole thermodynamics as
\begin{equation}
 d l^{2}=-\frac{1}{T}\left(\frac{\partial^{2}F}{\partial T^{2}}\right)_{V}d T^{2}
   +\frac{1}{T}\left(\frac{\partial^{2}F}{\partial V^{2}}\right)_{T}d V^{2}, \label{ddl2}
\end{equation}
where the free energy $F=U-TS$ and it satisfies the differential law $dF=-SdT-PdV$. By making use of the heat capacity at constant volume $C_{V}=T(\partial_{T}S)_{V}$, the line element reads
\begin{equation}
 d l^{2}=\frac{C_{V}}{T^{2}}d T^{2}-\frac{(\partial_{V}P)_{T}}{T}d V^{2}.\label{xxy}
\end{equation}
The scalar curvature for this geometry is \cite{WeiWeiWei}
\begin{eqnarray}
 R&=&\frac{1}{2C_{V}^{2}(\partial_{V}P)^{2}}
 \bigg\{
 T(\partial_{V}P)\bigg[(\partial_{T}C_{V})(\partial_{V}P-T\partial_{T,V}P)
 +(\partial_{V}C_{V})^{2}\bigg]\nonumber\\
 &+&C_{V}\bigg[(\partial_{V}P)^{2}+T\left((\partial_{V}C_{V})(\partial_{V,V}P)
 -T(\partial_{T,V}P)^{2}\right)
 +2T(\partial_{V}P)(T(\partial_{T,T,V}P)-(\partial_{V,V}C_{V}))\bigg]
 \bigg\}.\nonumber\\\label{RR}
\end{eqnarray}
Similar to the charged AdS black holes, here we also have $C_{V}$=0 for the charged GB-AdS black holes. This will lead to the divergence of $R$. In order to uncover the specific properties hiding behind it, we have defined a new normalized scalar curvature, which is
\begin{equation}
 R_{\rm N}=R*C_{V}=\frac{1}{2}-\frac{T^{2}(\partial_{V,T}P)^{2}-2T^{2}(\partial_{V}P)(\partial_{V,T,T}P)}{2(\partial_{V}P)^{2}}.
 \label{scalarcurv}
\end{equation}
Adopting this normalized scalar curvature, the microstructure has been explored for several different black holes.

\section{Phase diagram and microstructure in five dimensions}
\label{ptpsifd}

In this section, we shall focus on the five dimensions with $d$=5 and $\omega_{3}=2\pi^{2}$. In Refs. \cite{Wei2,Cai}, it was shown that there exists the small-large black hole phase transition. Here we shall examine its phase diagrams in detailed and construct the correspond Ruppeiner geometry.

\subsection{Phase diagrams and phase transitions}

\begin{figure}
\center{
\includegraphics[width=7cm]{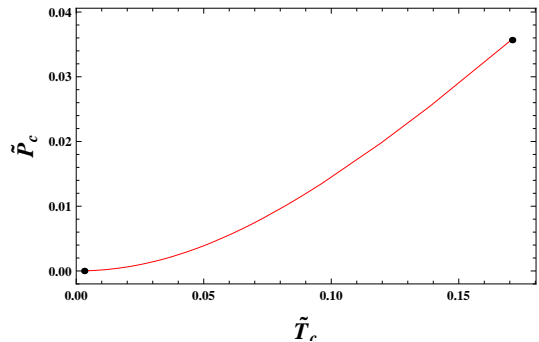}}
\caption{Critical points in the $\tilde{P}$-$\tilde{T}$ diagram. $\tilde{\alpha}$ increases from the right to left. The left and right dots are for $\tilde{\alpha}$=0 and $\infty$.}\label{pCP}
\end{figure}

For $d$=5, the reduced equation of state, the Gibbs free energy, and the heat capacity are given by
\begin{eqnarray}
 \tilde{P}&=&\frac{3\tilde{T}}{4\tilde{r}_{\text{h}}}-\frac{3}{8\pi \tilde{r}_{\text{h}}^{2}}+\frac{3\tilde{\alpha} \tilde{T}}{2\tilde{r}_{\text{h}}^{3}}+\frac{1}{8\pi \tilde{r}_{\text{h}}^{6}},\label{reeos}\\
 \tilde{G}&=&-\frac{\pi \left(2\pi \tilde{T} \tilde{r}_{\text{h}}^5-3\tilde{r}_{\text{h}}^4+36\pi \tilde{\alpha} \tilde{T}\tilde{r}_{\text{h}}^3-6 \tilde{\alpha}\tilde{r}_{\text{h}}^2-3\right)}{16\tilde{r}_{\text{h}}^2},\\
  C_{Q}/Q^{3/2}&=&\frac{3\pi^2\tilde{r}_{\text{h}}\left(8\pi P\tilde{r}_{\text{h}}^6+3 \tilde{r}_{\text{h}}^4-1\right) \left(2\tilde{\alpha}+\tilde{r}_{\text{h}}^2\right)^2}{2\left(6\tilde{\alpha}+3\tilde{r}_{\text{h}}^6(16\pi\tilde{\alpha}
   \tilde{P}-1)+8\pi \tilde{P}\tilde{r}_{\text{h}}^8+6\tilde{\alpha}\tilde{r}_{\text{h}}^4+5 \tilde{r}_{\text{h}}^2\right)}
\end{eqnarray}
The equation of state (\ref{reeos}) exhibits a small-large black hole phase transition, reminiscent of the liquid-gas phase transition of the VdW fluid. For the phase transition, there is a characteristic critical point indicating a second-order phase transition, which can be determined by
\begin{eqnarray}
 \left(\frac{\partial P}{\partial V}\right)_{T,Q,\alpha}=0,\quad
 \left(\frac{\partial^{2} P}{\partial V^{2}}\right)_{T,Q,\alpha}=0.
\end{eqnarray}
Since $V=\frac{\pi^{2}}{2}r_{\text{h}}^{4}$ for $d$=5, the conditions can be expressed as
\begin{eqnarray}
 \left(\frac{\partial \tilde{P}}{\partial \tilde{r}_{\text{h}}}\right)_{\tilde{T},\tilde{\alpha}}=0,\quad
 \left(\frac{\partial^{2} \tilde{P}}{\partial \tilde{r}_{\text{h}}^{2}}\right)_{\tilde{T},\tilde{\alpha}}=0.
\end{eqnarray}
Solving the equation, we have
\begin{eqnarray}
 4 \tilde{r}_{\text{h}}^6-24\tilde{\alpha} \tilde{r}_{\text{h}}^4-5\tilde{r}_{\text{h}}^2-18\tilde{\alpha}=0.
\end{eqnarray}
Then the critical value of the radius of the horizon can be obtained
\begin{eqnarray}
 \tilde{r}_{\text{hc}}&=&\sqrt{2\alpha+2\sqrt{\frac{5+12\tilde{\alpha}^2}{3}}\cos\Theta},\\
 \Theta&=&\frac{1}{3}\arccos\left(\frac{6\sqrt{3}\tilde{\alpha}(7+4\tilde{\alpha}^2)}{(5+12\tilde{\alpha}^2)^{\frac{3}{2}}}\right).
\end{eqnarray}
The corresponding critical pressure and temperature are
\begin{eqnarray}
 \tilde{P}_{\text{c}}&=&-\frac{6 \tilde{\alpha} +6 \tilde{\alpha}
   \tilde{r}_{\text{hc}}^4-3 \tilde{r}_{\text{hc}}^6+5 \tilde{r}_{\text{hc}}^2}{48
   \pi  \tilde{\alpha}  \tilde{r}_{\text{hc}}^6+8 \pi \tilde{r}_{\text{hc}}^8},\\
 \tilde{T}_{\text{c}}&=&\frac{\tilde{r}_{\text{hc}}^4-1}{\pi \tilde{r}_{\text{hc}}^3
   \left(6 \tilde{\alpha} +\tilde{r}_{\text{hc}}^2\right)}.
\end{eqnarray}
We plot the critical point in the $\tilde{P}$-$\tilde{T}$ diagram in Fig. \ref{pCP} for varying $\tilde{\alpha}$. When $\tilde{\alpha}$=0, the critical point curve starts at ($\frac{4}{5\sqrt[4]{5}\pi}$, $\frac{1}{4\sqrt{5}\pi}$). Then with the increase of $\tilde{\alpha}$, the curve extends to low temperature and pressure, and ends at (0, 0) when $\tilde{\alpha}\rightarrow\infty$. Each point on this curve is a phase transition point of second-order.

\begin{figure}
\center{\subfigure[]{\label{Sheat04a}
\includegraphics[width=7cm]{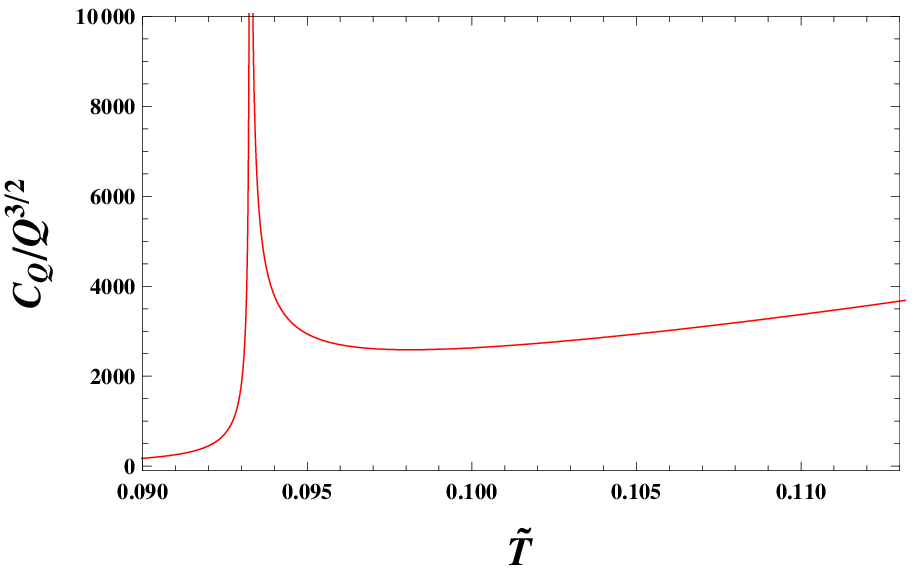}}
\subfigure[]{\label{Sheat14b}
\includegraphics[width=7cm]{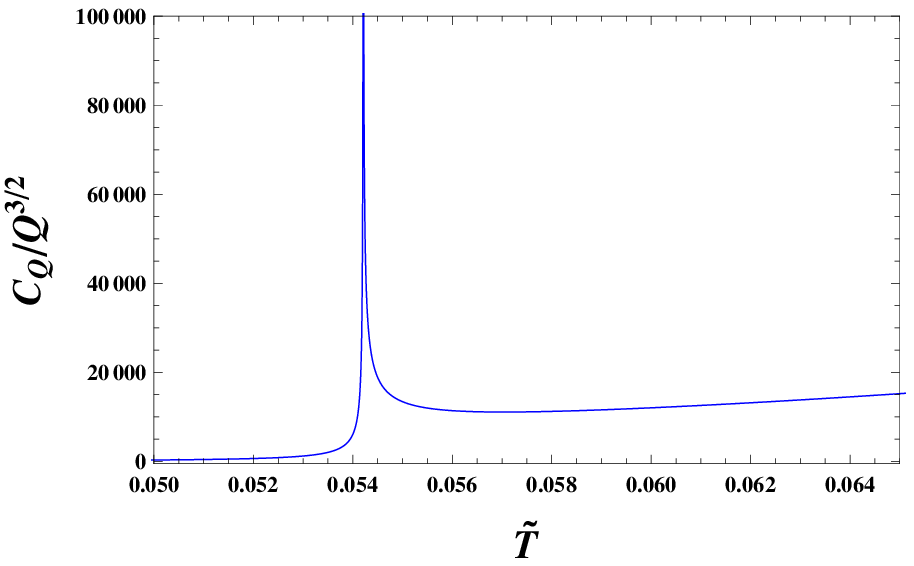}}}
\caption{Heat capacity as a function of the temperature at fixed critical pressure. The divergent behaviors occur at the critical temperature. (a) $\tilde{\alpha}$=0.4. (b) $\tilde{\alpha}$=1.4. Obviously, the $\lambda$ phase transition behavior is observed.}\label{pSheat14b}
\end{figure}

For $\tilde{\alpha}$=0.4 and 1.4, we plot the heat capacity in Fig. \ref{pSheat14b} when the pressure takes its critical values. Obviously, we observe the $\lambda$ behaviors for these two cases. Such pattern of the heat capacity is reminiscent of that of $^4$He. For black hole systems, this behavior was first observed in Ref. \cite{Tjoa2}, where a new superfluid was uncovered. Here, we suggest that such $\lambda$ behavior of the heat capacity implies a second-order phase transition rather a new black hole phase unless the conditions given in Ref. \cite{Tjoa2} are satisfied.

Employing these critical thermodynamic quantities, we can construct a universal quantity
\begin{eqnarray}
 \rho=\frac{\tilde{P}_{\text{c}}\tilde{r}_{\text{hc}}}{\tilde{T}_{\text{c}}}
 =\frac{6\tilde{\alpha}-3\tilde{r}_{\text{hc}}^6+6\tilde{\alpha}\tilde{r}_{\text{hc}}^4+5
   \tilde{r}_{\text{hc}}^2}{8\tilde{r}_{\text{hc}}^2(1-\tilde{r}_{\text{hc}}^4)}.
\end{eqnarray}
For $\alpha$=0, we have $\rho$=5/16.

\begin{figure}
\center{\subfigure[]{\label{ISOtemp04a}
\includegraphics[width=7cm]{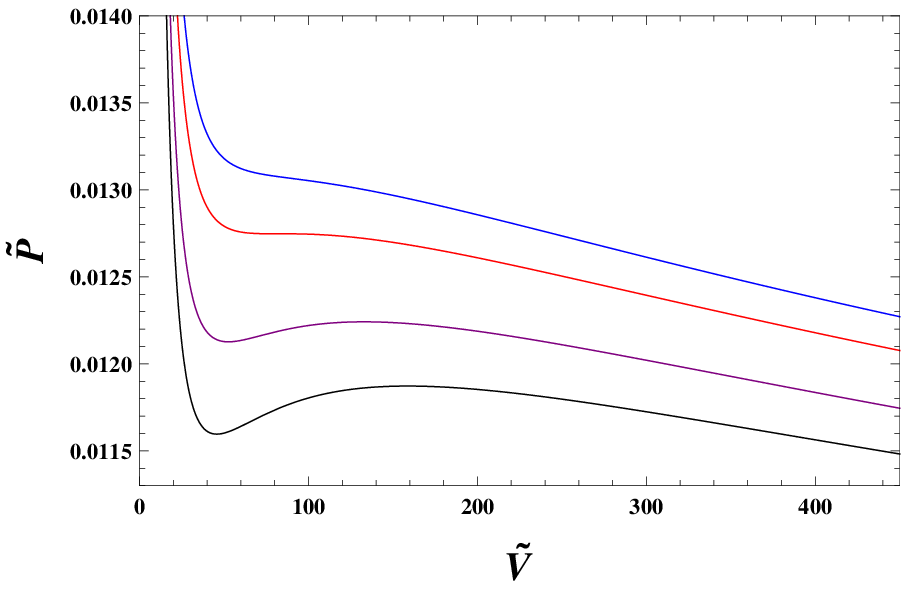}}
\subfigure[]{\label{ISOtemp14b}
\includegraphics[width=7cm]{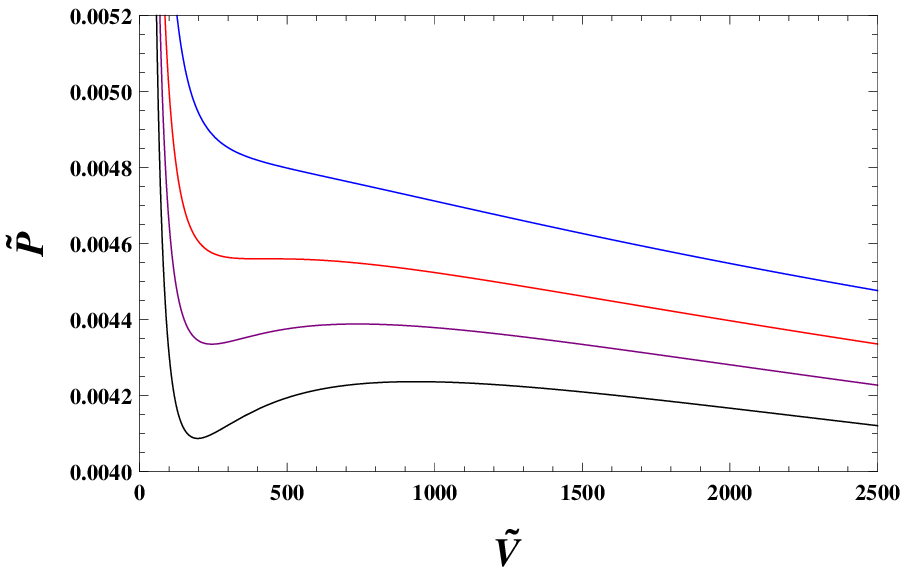}}}
\caption{The isothermal curves for the black hole in the $\tilde{P}$-$\tilde{V}$ diagram for (a) $\tilde{\alpha}$=0.4. $\tilde{T}$=0.091, 0.092, 0.093261 ($\tilde{T}_{\text{c}}$), and 0.094 from bottom to top. (b) $\tilde{\alpha}$=1.4. $\tilde{T}$=0.053, 0.0536, 0.054209 ($\tilde{T}_{\text{c}}$), and 0.055 from bottom to top.}\label{pISOtemp14b}
\end{figure}

Next, we plot the isothermal curves and the Gibbs free energies for further understanding the black hole phase transition. The isothermal curves are shown in Fig. \ref{pISOtemp14b} in the $\tilde{P}$-$\tilde{V}$ diagram for $\tilde{\alpha}$=0.4 and 1.4, respectively. From these figures, one can observe nonmonotonic behavior for low temperature, which indicates the existence of the small-large phase transition. For fixed temperature, these black hole branches of the positive slope are nonphysical. They should be removed by other mechanisms, one of which is the Maxwell equal area law. One can draw a horizontal line to construct two equal areas on one isothermal curve. Then the pressure and temperature corresponding to the horizontal line denote the values of the phase transition point. In our reduced treatment, it is equivalent to fix the charge, so the equal area law is also applicable in the $\tilde{P}$-$\tilde{V}$ diagram on determining the phase transition point.

\begin{figure}
\center{\subfigure[]{\label{FiveGIbbs04a}
\includegraphics[width=7cm]{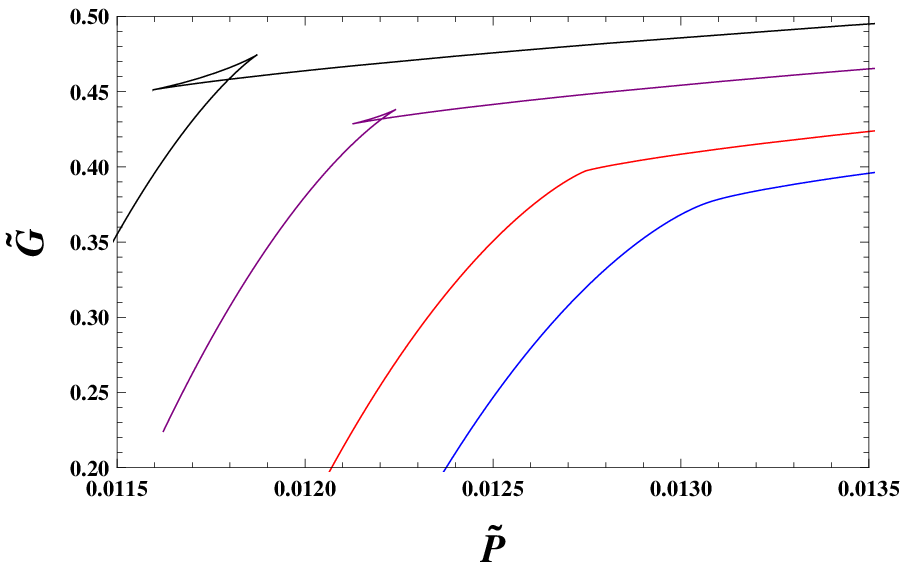}}
\subfigure[]{\label{FiveGIbbs14b}
\includegraphics[width=7cm]{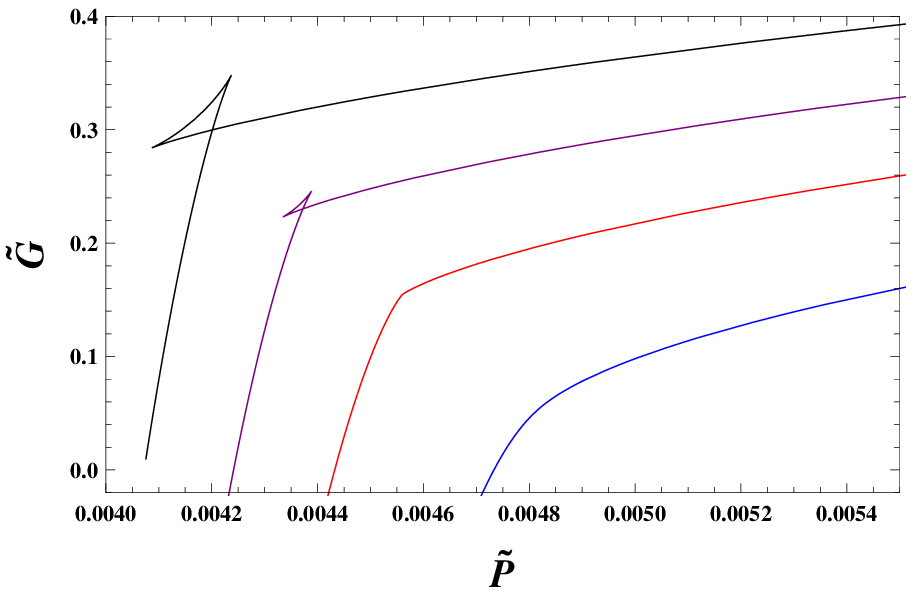}}}
\caption{Behaviors of the Gibbs free energy for fixed temperature. a) $\tilde{\alpha}$=0.4. $\tilde{T}$=0.091, 0.092, 0.093261 ($\tilde{T}_{\text{c}}$), and 0.094 from left to right. (b) $\tilde{\alpha}$=1.4. $\tilde{T}$=$\tilde{T}$=0.053, 0.0536, 0.054209 ($\tilde{T}_{\text{c}}$), and 0.055 from left to right.}\label{pFiveGIbbs14b}
\end{figure}

Another alternative on determining the phase transition is the Gibbs free energy. We show its behavior in Fig. \ref{pFiveGIbbs14b}. When the temperature is lower than the critical value, there is the characteristic swallow tail behavior. Combining the thermodynamic choice that a system always prefers the lowest free energy, the phase transition exactly occurs at the intersection point of the Gibbs free energy. Therefore, one can also use such property to numerically solve the phase transition point.

\begin{figure}
\center{\subfigure[]{\label{PTPhaseDa04}
\includegraphics[width=7cm]{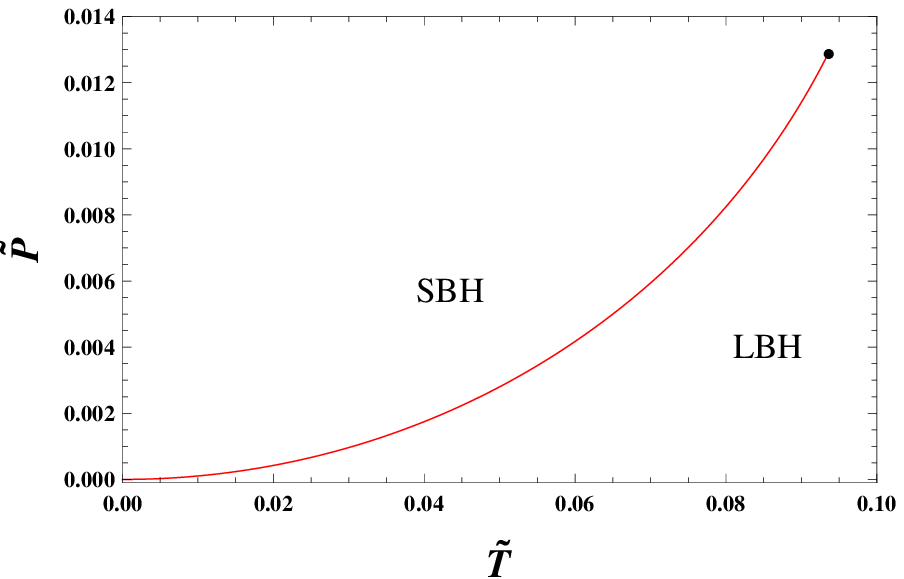}}
\subfigure[]{\label{PTPhaseDb14}
\includegraphics[width=7cm]{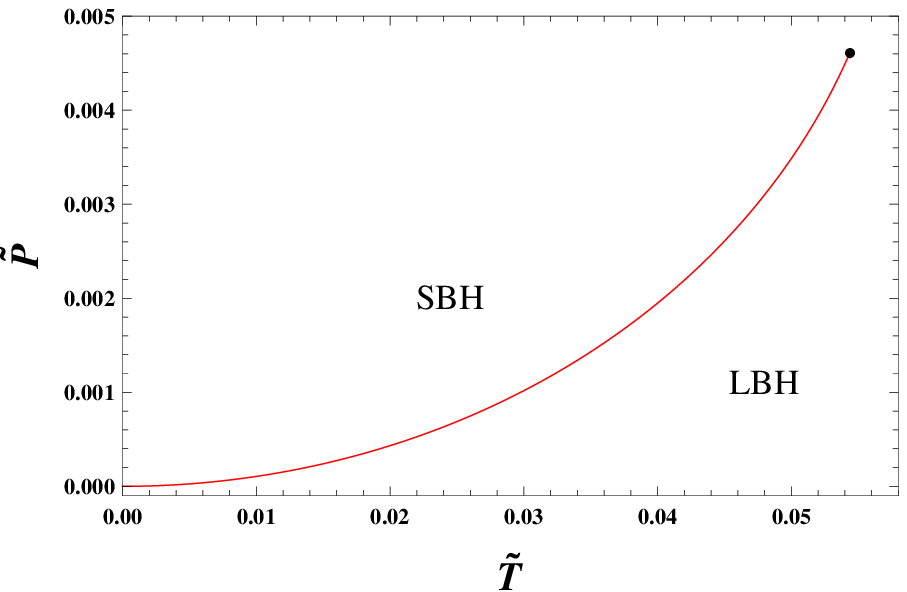}}}
\caption{Phase structures in $\tilde{P}$-$\tilde{T}$ diagram. (a) $\tilde{\alpha}$=0.4. (b) $\tilde{\alpha}$=1.4. SBH and LBH
are for the small and large black holes. Black dots denote the critical points.}\label{pPTPhaseDb14}
\end{figure}

\begin{figure}
\center{\subfigure[]{\label{TrhPhaseDa04}
\includegraphics[width=7cm]{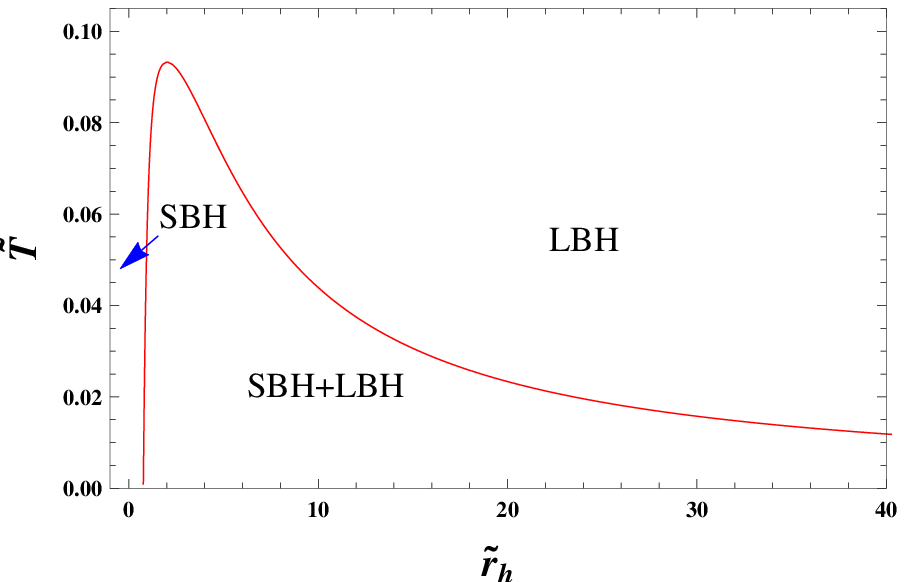}}
\subfigure[]{\label{TrhPhaseDb14}
\includegraphics[width=7cm]{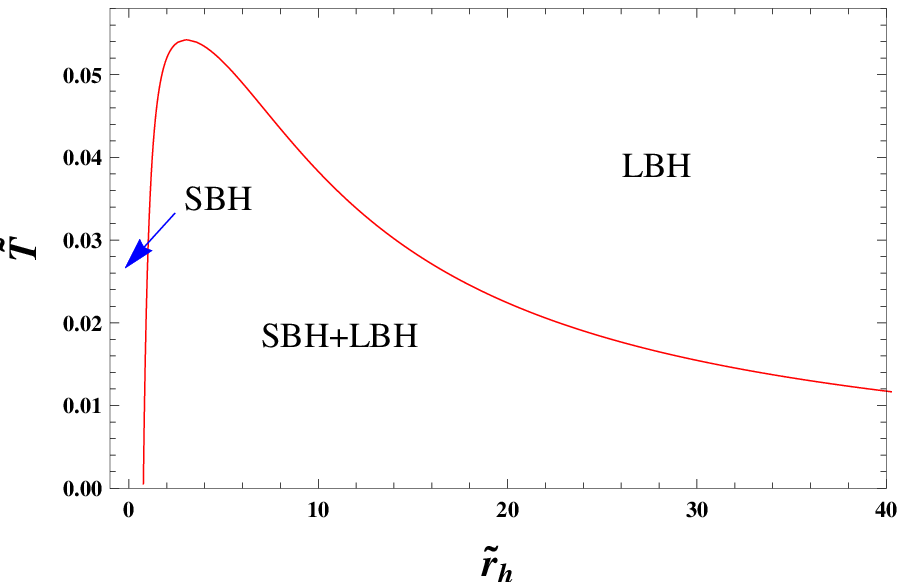}}}
\caption{Phase structures in the $\tilde{T}$-$\tilde{r}_{\text{h}}$ diagram. (a) $\tilde{\alpha}$=0.4. (b) $\tilde{\alpha}$=1.4.}\label{pTrhPhaseDb14}
\end{figure}

Actually, these two approaches coincide with each other when the first law of the black hole thermodynamics holds. Here we display the phase structures in the $\tilde{P}$-$\tilde{T}$ and $\tilde{T}$-$\tilde{r}_{\text{h}}$ diagrams in Figs. \ref{pPTPhaseDb14} and \ref{pTrhPhaseDb14}, respectively. In the $\tilde{P}$-$\tilde{T}$ diagram shown in Fig. \ref{pPTPhaseDb14}, the red solid curves are for the coexistence phase of small and large black holes. Above the curve is the small black hole region and below it is the large black hole region. Black dots denote the critical points, where the second-order phase transition occurs. Moreover, we also observe that the critical point is shifted toward low temperature and pressure for big $\tilde{\alpha}$. In Fig. \ref{pTrhPhaseDb14}, the phase diagram is described in the $\tilde{T}$-$\tilde{r}_{\text{h}}$ diagram. Obviously, the coexistence curve given in Fig. \ref{pPTPhaseDb14} becomes the coexistence region, see below the red curve. The small and large black hole regions, respectively, locate at the left and right. For different $\tilde{\alpha}$, the pattern is similar, while with different peaks indicating different critical points.

\begin{figure}
\center{\subfigure[]{\label{DerhLa}
\includegraphics[width=7cm]{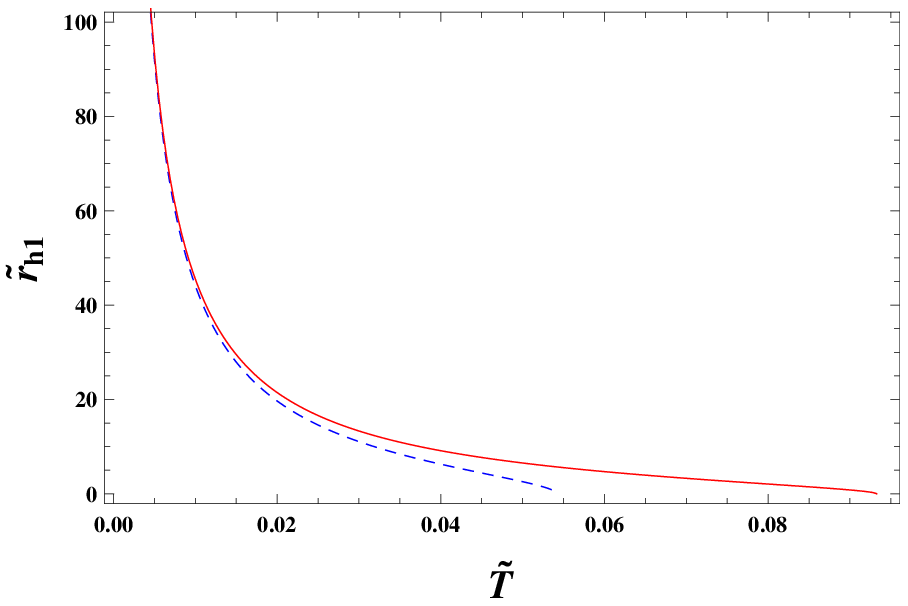}}
\subfigure[]{\label{DerhSb}
\includegraphics[width=7cm]{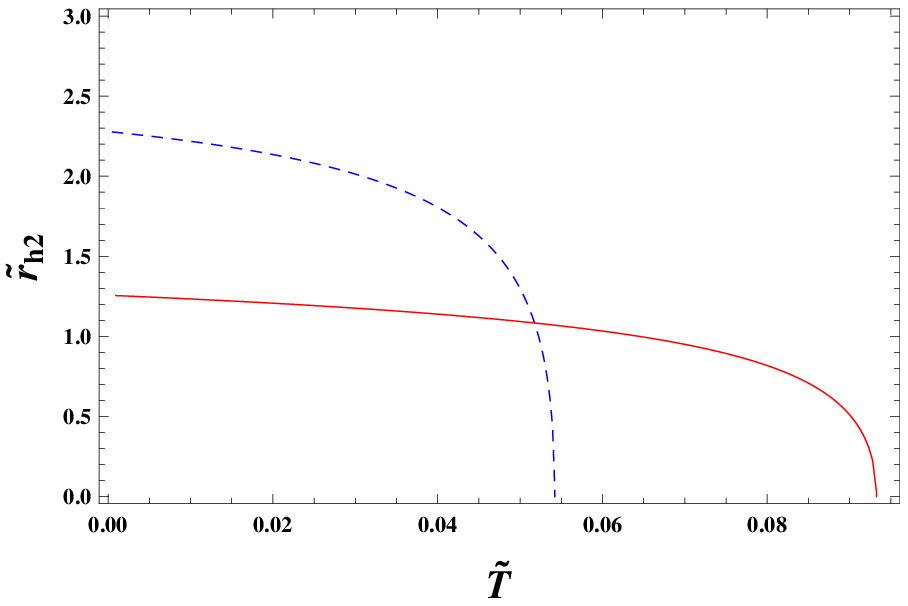}}}
\caption{Behaviors of $\tilde{r}_{\text{h1}}$ (a) and $\tilde{r}_{\text{h2}}$ (b). Red solid curves are for $\tilde{\alpha}$=0.4 and blue dashed curves are for $\tilde{\alpha}$=1.4.}\label{pDerhSb}
\end{figure}

As shown in Fig. \ref{pTrhPhaseDb14}, the radius of the black hole horizon has a sudden change for a fixing phase transition temperature, which indicates that black hole microstructure gets a significant change. Now we define two new radii shifted by the critical value of the horizon radius at the phase transition
\begin{eqnarray}
 \tilde{r}_{\text{h1}}=\tilde{r}_{\text{hl}}-\tilde{r}_{\text{hc}},\quad
 \tilde{r}_{\text{h2}}=\tilde{r}_{\text{hc}}-\tilde{r}_{\text{hs}},
\end{eqnarray}
where $\tilde{r}_{\text{hs}}$ and $\tilde{r}_{\text{hl}}$ denote the horizon radius of the coexistence small and large black holes. We describe $\tilde{r}_{\text{h1}}$ and $\tilde{r}_{\text{h2}}$ in Fig. \ref{pDerhSb}. From Fig. \ref{DerhLa}, we find that $\tilde{r}_{\text{h1}}$ has large values near $\tilde{T}$=0. Then it decreases with $\tilde{T}$. At the critical temperature, $\tilde{r}_{\text{h1}}$ vanishes for $\tilde{\alpha}$=0.4 and 1.4. Similar phenomena can also be observed for $\tilde{r}_{\text{h2}}$, see in Fig. \ref{DerhSb}. The only difference is that near $\tilde{T}$=0, $\tilde{r}_{\text{h2}}$ is finite. For example, $\tilde{r}_{\text{h2}}$=1.25 and 2.28 for $\tilde{\alpha}$=0.4 and 1.4, respectively. The reason is that $\tilde{r}_{\text{hl}}$ has no boundary, while $\tilde{r}_{\text{hs}}$ cannot be smaller than zero.

As shown above $\tilde{r}_{\text{h1}}$ and $\tilde{r}_{\text{h2}}$ vanish at the critical temperature, now, we aim to calculate its critical exponent. After the numerical calculation near $\tilde{T}$=0, we fit the data with the following form
\begin{eqnarray}
 \tilde{r}_{\text{h1,2}}=a(\tilde{T}_{\text{c}}-\tilde{T})^{b},\label{fit1}
\end{eqnarray}
where $a$ and $b$ are the fitting coefficients. We show the critical points and fitting coefficients in Table \ref{parameters}. Focusing on the coefficient $b$, we find that all them are near 1/2 with the relative deviation within 1.6\%. Actually this value can also be obtained via expanding the equation of state of the system near the critical point. So we can conclude that $\tilde{r}_{\text{h1}}$ and $\tilde{r}_{\text{h2}}$ have a critical exponent 1/2. Hence $\tilde{r}_{h1,2}$, or the horizon radius, can serve as order parameters to characterize the small-large black hole phase transition. This also provides a fundamental interpretation of the horizon radius on investigating the dynamic process of the phase transitions.

\begin{table}[h]
\begin{center}
\begin{tabular}{c c c c c c }
  \hline
  \hline
   $\alpha$&$\tilde{P}_{\text{c}}$ &$\tilde{T}_{\text{c}}$ & $\tilde{r}_{\text{hc}}$ & $a$ & $b$\\
  \hline
 0.4 & 0.012748 & 0.093261 & 2.016305  & 12.8281/10.0179 & 0.5095/0.4903  \\
 1.4 & 0.004560 & 0.054209 & 3.039192  & 32.6572/26.0413 & 0.5078/0.4913  \\
  \hline\hline
\end{tabular}
\end{center}
\caption{Values of the critical points and fitting coefficients for $\tilde{\alpha}$=0.4 and $\tilde{\alpha}$=1.4. .../... means the data are for $\tilde{r}_{\text{h1}}$/$\tilde{r}_{\text{h2}}$.}\label{parameters}
\end{table}

\subsection{Ruppeiner geometry and microstructure}

As observed in Refs. \cite{Wei2020a,Zhourun}, the neutral GB-AdS black hole has an intriguing property that its interaction among the microscopic constituents maintains the same while its microstructure behaves different before and after the small-large black hole phase transition. In this subsection, we shall study the microstructure of the black hole in five dimensions and see whether this property holds for the charged case in the canonical ensemble.

\begin{figure}
\center{
\includegraphics[width=7cm]{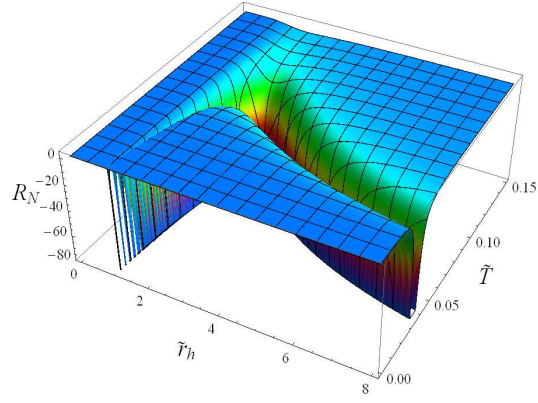}}
\caption{The normalized scalar curvature $R_{\text{N}}$ as a function of $\tilde{T}$ and $\tilde{r}_{\text{h}}$ for $\tilde{\alpha}=0.4$.}\label{RTRH}
\end{figure}

Making use of the equation of state (\ref{reeos}), we obtain the corresponding normalized scalar curvature of the Ruppeiner geometry via (\ref{scalarcurv}). When expressing in terms of $\tilde{T}$ and $\tilde{r}_{\text{h}}$, we have
\begin{eqnarray}
 R_{\text{N}}=-\frac{\left(4\tilde{r}_{\text{h}}^4-1\right)\left(12\pi\tilde{\alpha} \tilde{T} \tilde{r}_{\text{h}}^3+2\pi \tilde{T} \tilde{r}_{\text{h}}^5- \tilde{r}_{\text{h}}^4+1\right)}{2\left(6\pi \tilde{\alpha} \tilde{T} \tilde{r}_{\text{h}}^3+\pi \tilde{T}\tilde{r}_{\text{h}}^5- \tilde{r}_{\text{h}}^4+1\right)^2}.\label{rn5}
\end{eqnarray}
Obviously, the charge is absent by using the reduced thermodynamic quantities. We plot $R_{\text{N}}$ for different $\tilde{T}$ and $\tilde{r}_{\text{h}}$ for $\tilde{\alpha}=0.4$ in Fig. \ref{RTRH}. Other values of $\tilde{\alpha}$ share the similar pattern. For low temperature, $R_{\text{N}}$ has two negative divergences at two $\tilde{r}_{\text{h}}$. These two divergences merger at the critical point, and disappear beyond the critical temperature.

\begin{figure}
\center{\subfigure[]{\label{RNrha}
\includegraphics[width=7cm]{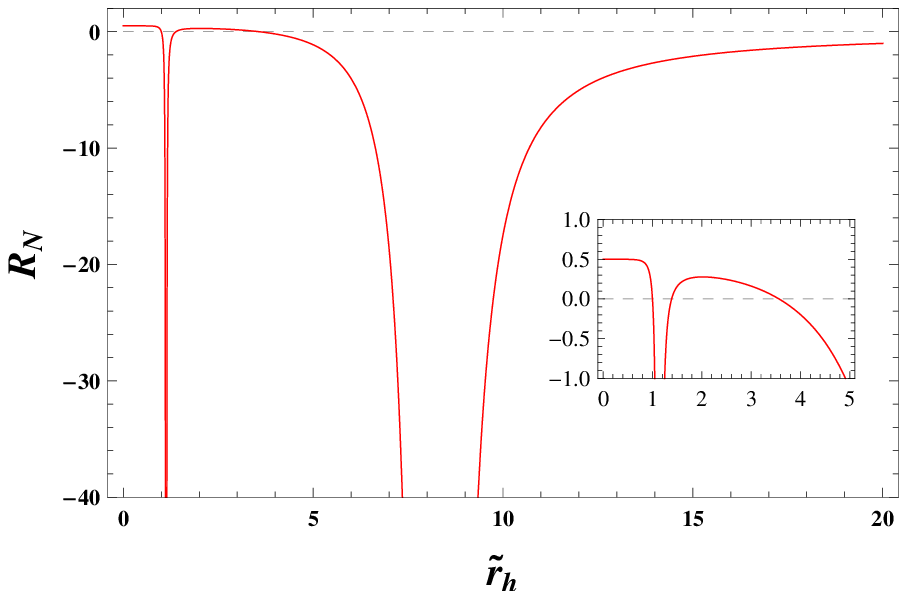}}
\subfigure[]{\label{RNrhb}
\includegraphics[width=7cm]{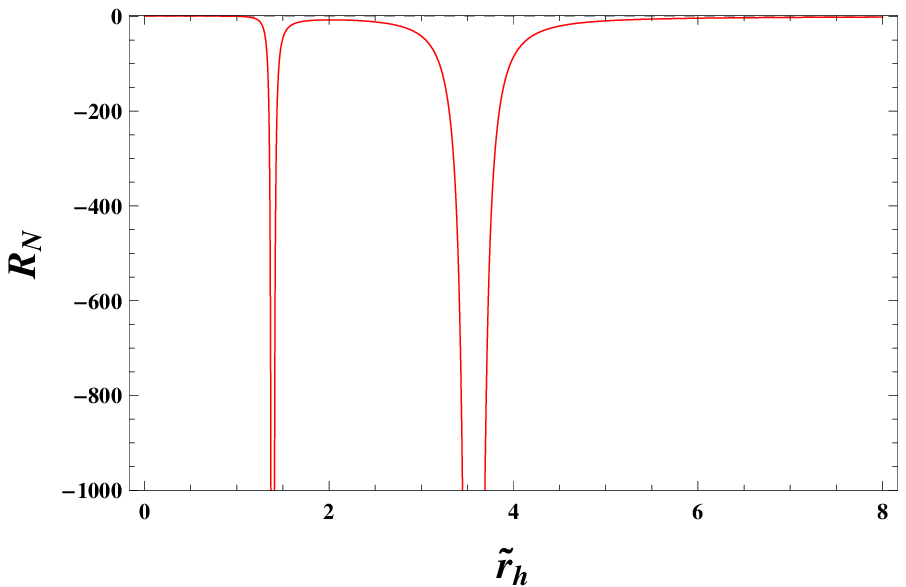}}\\
\subfigure[]{\label{RNrhc}
\includegraphics[width=7cm]{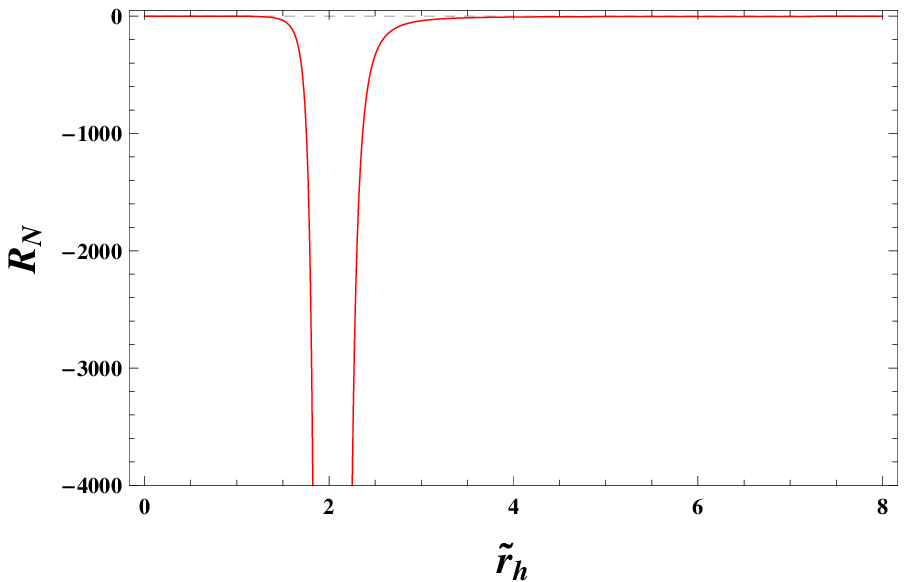}}
\subfigure[]{\label{RNrhd}
\includegraphics[width=7cm]{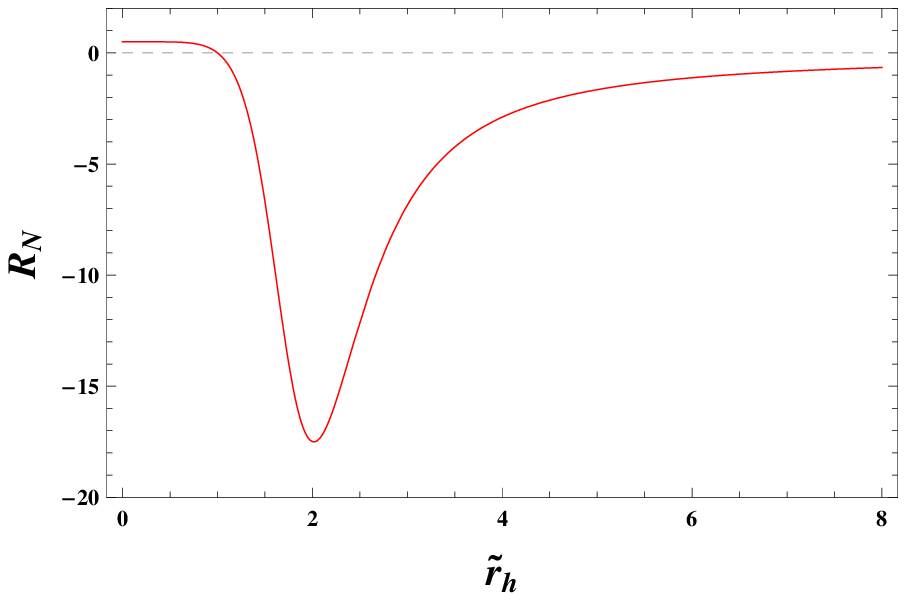}}}
\caption{Behaviors of $R_{\text{N}}$ for different fixing temperature with $\tilde{\alpha}$=0.4. (a) $\tilde{T}$=0.4$\tilde{T}_{\text{c}}$. (b) $\tilde{T}$=0.8$\tilde{T}_{\text{c}}$. (c) $\tilde{T}$=$\tilde{T}_{\text{c}}$. (d) $\tilde{T}$=1.2$\tilde{T}_{\text{c}}$. }\label{pRNrhd}
\end{figure}

In order to show the details, we display $R_{\text{N}}$ for $\tilde{T}$=0.4$\tilde{T}_{\text{c}}$, 0.8$\tilde{T}_{\text{c}}$, $\tilde{T}_{\text{c}}$, and 1.2 $\tilde{T}_{\text{c}}$ in Fig. \ref{pRNrhd}. It is clear that there are two negative divergent points for $\tilde{T}$=0.4$\tilde{T}_{\text{c}}$ and 0.8$\tilde{T}_{\text{c}}$, and one for $\tilde{T}=\tilde{T}_{\text{c}}$. While when $\tilde{T}=1.2\tilde{T}_{\text{c}}$, there is only a well with finite depth. More importantly, for $\tilde{T}$=0.4$\tilde{T}_{\text{c}}$ shown in Fig. \ref{RNrha}, we find that there are two parameter regions, where $R_{\text{N}}$ is positive. One of them is at small $\tilde{r}_{\text{h}}$. For different temperatures, we always observe positive $R_{\text{N}}$. Another region is near $\tilde{r}_{\text{h}}$=2. However, it only appears for $\tilde{T}$=0.4$\tilde{T}_{\text{c}}$.

Now, we aim to uncover the divergent point and zero point of $R_{\text{N}}$. Solving (\ref{rn5}), we obtain the divergent temperature $\tilde{T}_{\text{sp}}$ of $R_{\text{N}}$, which gives
\begin{eqnarray}
 \tilde{T}_{\text{sp}}=\frac{\tilde{r}_{\text{h}}^4-1}{\pi\tilde{r}_{\text{h}}^3(\tilde{r}_{\text{h}}^2+6\tilde{\alpha})}.
\end{eqnarray}
Actually, the curve described by this temperature separates the metastable and unstable black hole branches, so we can call it the spinodal curve.
The temperature corresponding to zero point is
\begin{eqnarray}
 \tilde{T}_{0}=\frac{1}{2}\tilde{T}_{\text{sp}}.
\end{eqnarray}
Besides, at
\begin{eqnarray}
 \tilde{r}_{\text{h}}=1,
\end{eqnarray}
$R_{\text{N}}$ will also vanish. We name the zero point curve of $R_{\text{N}}$ as the sign-changing point, across which $R_{\text{N}}$ changes its sign.

\begin{figure}
\center{\subfigure[]{\label{RTrhPhasea}
\includegraphics[width=7cm]{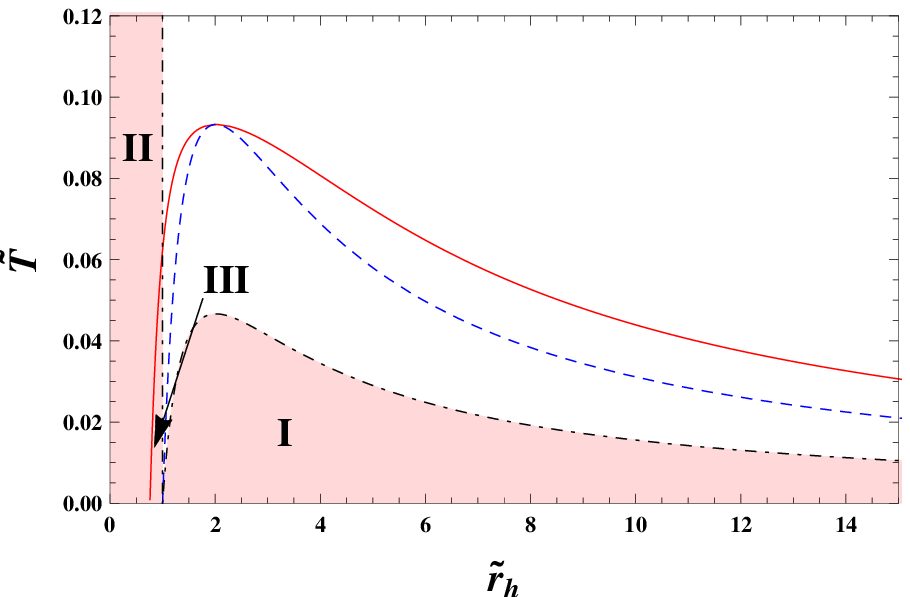}}
\subfigure[]{\label{RTrhPhaseb}
\includegraphics[width=7cm]{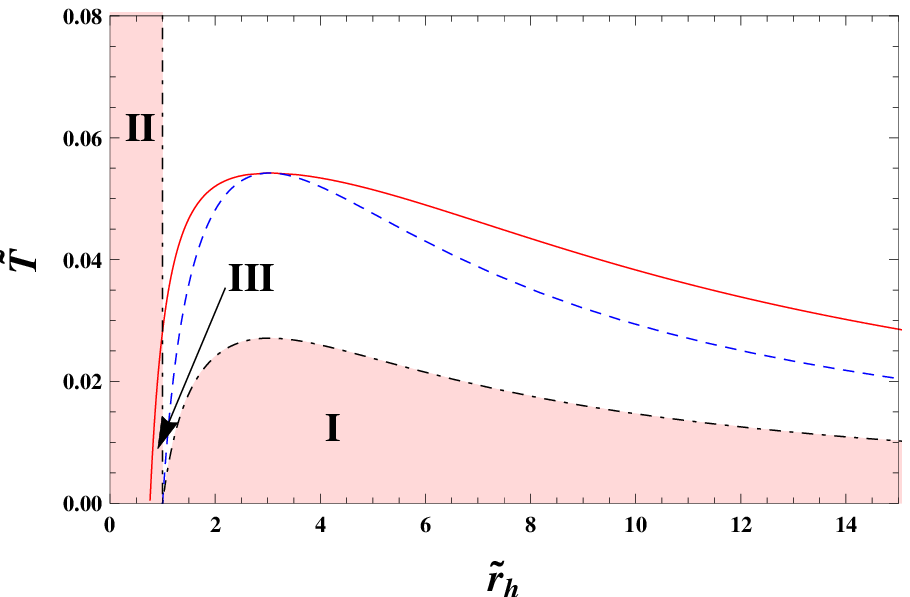}}}
\caption{The coexistence curves (red solid lines), spinodal curves (blue dashed lines), and the sign-changing curves of $R_{\text{N}}$ (black dot-dashed lines). The normalized scalar curvature $R_{\text{N}}$ diverges at the spinodal curves. Both the spinodal and sign-changing curves start at $\tilde{r}_{\text{h}}$=$1/\sqrt{2}$. In the shadow regions I, II, and III, $R_{\text{N}}$ is positive, otherwise, it is negative. (a) $\tilde{\alpha}$=0.4. (b) $\tilde{\alpha}$=1.4.}\label{pRTrhPhaseb}
\end{figure}

We describe these characteristic curves in the phase diagram in Fig. \ref{pRTrhPhaseb} for $\tilde{\alpha}$=0.4 and 1.4, respectively. The red solid curves are the coexistence curves of small and large black holes, below which is the coexistence regions, and the equation of state (\ref{reeos}) may not hold. So we need to exclude such regions when we consider the normalized scalar curvature $R_{\text{N}}$, where we have used the equation of state in the calculation. The spinodal curves are marked in blue dashed lines. They only meet the coexistence curves at the critical points. The black dot-dashed lines are for the sign-changing curves, which separate the phase diagram into two regions, one with positive $R_{\text{N}}$ and another with negative $R_{\text{N}}$ shown in shadow regions.

For each $\tilde{\alpha}$, we find that there are three regions I, II, and III that have positive $R_{\text{N}}$, indicating that the repulsive interaction dominates among the black hole microstructure. While in other regions, the attractive interaction dominates. Since regions I and III are in the coexistence regions, we exclude them in our consideration. However, in region II, the small black holes with higher temperature always have positive $R_{\text{N}}$. So the repulsive interaction can be observed for the charged GB-AdS black holes. This behavior is quite similar to the charged AdS black hole without the GB coupling \cite{WeiWeiWei}, while different from the neutral GB-AdS black holes \cite{Wei2020a,Zhourun}.

\begin{figure}
\center{\subfigure[]{\label{RNT04a}
\includegraphics[width=7cm]{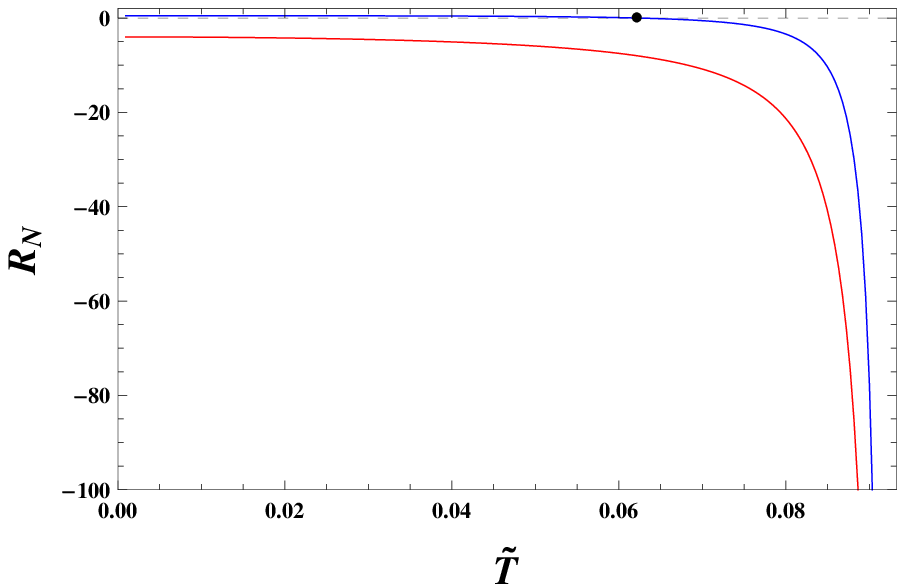}}
\subfigure[]{\label{RNT14b}
\includegraphics[width=7cm]{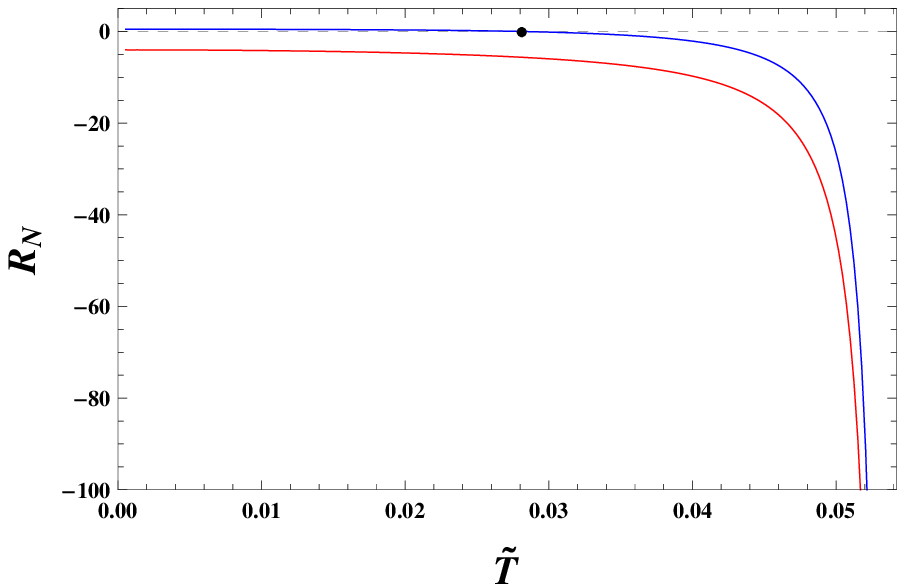}}}
\caption{$R_{\text{N}}$ along the coexistence small (top blue curves) and large (bottom red curves) black holes. Black dots denote the zero points of $R_{\text{N}}$, which are at $\tilde{T}$=0.062312 and 0.028235, respectively. (a) $\tilde{\alpha}$=0.4. (b) $\tilde{\alpha}$=1.4.}\label{pRNT14b}
\end{figure}

We also show the behaviors of $R_{\text{N}}$ along the coexistence small and large black hole curves for $\tilde{\alpha}$=0.4 and 1.4 in Fig \ref{pRNT14b}. With the increase of the temperature, both the curves decrease, and go to negative infinity at the critical temperature. In particular, $R_{\text{N}}$ is positive at low temperature, $T<$0.062312 for $\tilde{\alpha}$=0.4 and $T<$0.028235 for $\tilde{\alpha}$=1.4 when it goes along the coexistence small black holes indicating that the repulsive interaction exists.

\begin{figure}
\center{\subfigure[]{\label{LNRTa}
\includegraphics[width=7cm]{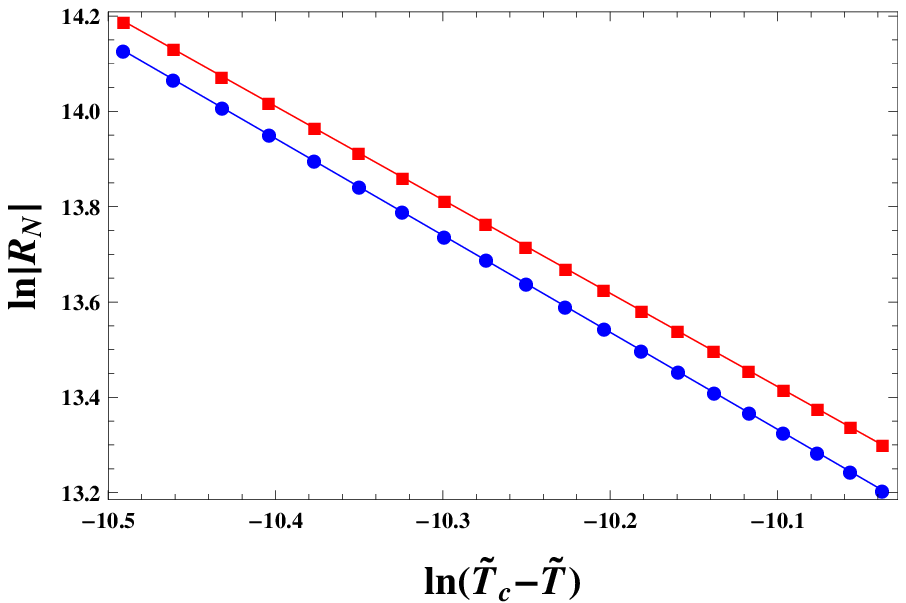}}
\subfigure[]{\label{LNRTb}
\includegraphics[width=7cm]{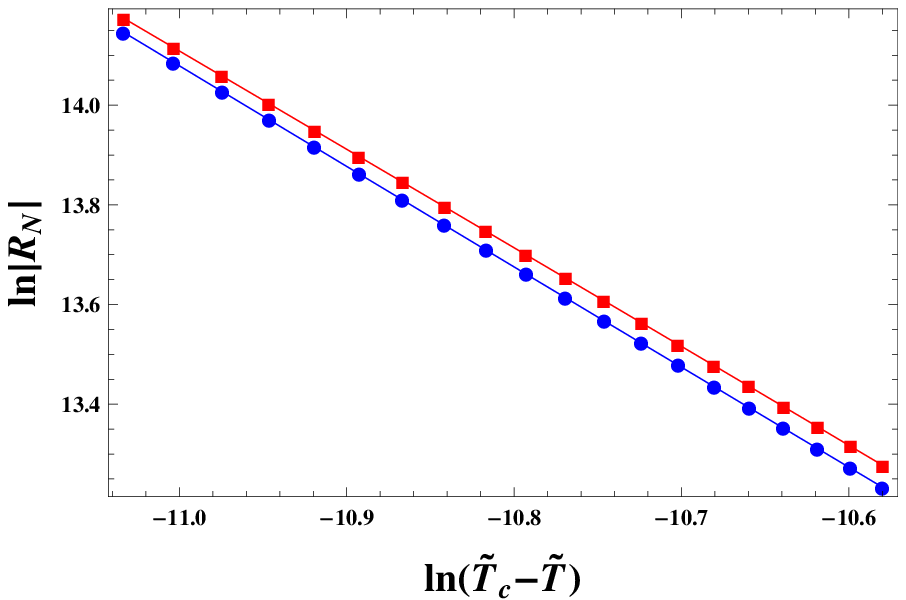}}}
\caption{$\ln|R_{\text{N}}|$ vs. $\ln(\tilde{T}_{\text{c}}-\tilde{T})$ when the temperature is very near its critical values. Bottom blue and top red curves are for the coexistence small and large black holes. (a) $\tilde{\alpha}$=0.4.  (b) $\tilde{\alpha}$=1.4.}\label{pLNRTb}
\end{figure}

At the critical temperature, all these curves go to negative infinity. Here we plot $\ln|R_{\text{N}}|$ as a function of $\ln(\tilde{T}_{\text{c}}-\tilde{T})$ in Fig. \ref{pLNRTb} near the critical temperature. An intuitive impression is that it has a linear relation. We fit the data in the following form
\begin{eqnarray}
 \ln|R_{\text{N}}|=\beta\ln(\tilde{T}_{\text{c}}-\tilde{T})+\gamma.\label{fit2}
\end{eqnarray}
The fitting coefficients are listed in Table \ref{par2}. From the fitting results, we observe that the coefficient $\beta$ is much near -2 of relative deviation within $1\%$, which suggests
\begin{eqnarray}
 R_{\text{N}}\sim (\tilde{T}_{\text{c}}-\tilde{T})^{-2}.
\end{eqnarray}
So $R_{\text{N}}$ has a critical exponent 2. Actually, this critical exponent can also be obtained by expanding the equation of state near the critical point.

\begin{table}[h]
\begin{center}
\begin{tabular}{c c c c c}
  \hline
  \hline
   $\alpha$& $\beta$(SBH) & $\beta$(LBH) & $\gamma$(SBH) & $\gamma$(LBH)\\
  \hline
 0.4 & -2.0342 &-1.9609 & -7.2129  & -6.3822   \\
 1.4 & -2.0131 &-1.9774 & -8.0660  & -7.6425   \\
  \hline\hline
\end{tabular}
\end{center}
\caption{Fitting coefficients $\beta$ and $\gamma$ along the coexistence small and large black holes.}\label{par2}
\end{table}

In summary, we studied the thermodynamics and Ruppeiner geometry for the five-dimensional charged AdS black holes. Different from the neutral case, we observe that the repulsive interaction could exist for the small black hole with high temperature. Meanwhile, along the coexistence small or large black hole curve, $R_{\text{N}}$ takes different values indicating that the microscopic interaction changes among the black hole phase transition, where the microstructure changes either. The similar thing is that the normalized scalar curvature goes to negative infinity at the critical point, and it has a critical exponent 2.

\section{Phase diagram and microstructure in more than five dimensions}
\label{ptinfd}

Next, we consider the case with the dimension number $d>5$ and confirm whether the similar properties of $d$=5 hold.

\begin{figure}
\center{\subfigure[]{\label{CTd6a}
\includegraphics[width=7cm]{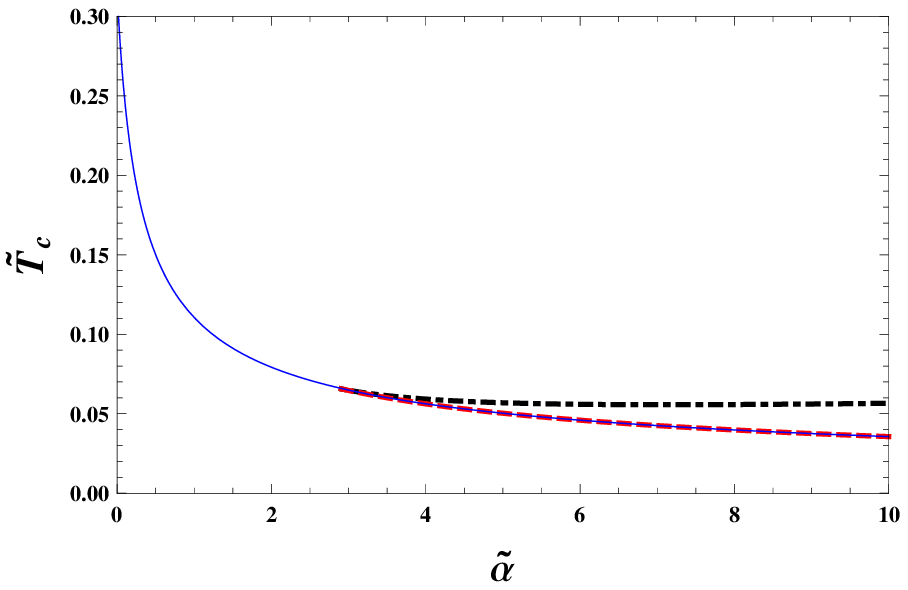}}
\subfigure[]{\label{CTd6b}
\includegraphics[width=7cm]{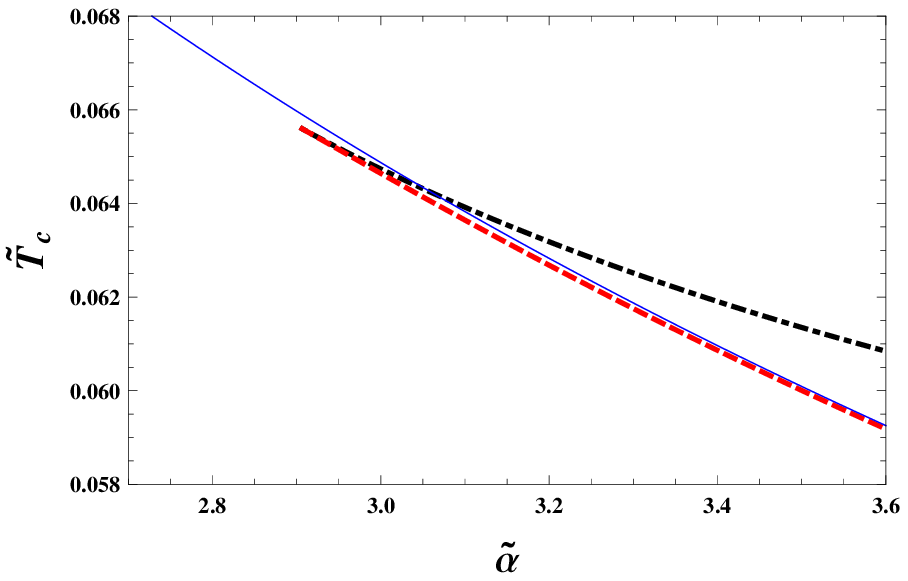}}}
\caption{Critical temperature $\tilde{T}_{\text{c}}$ as a function of $\tilde{\alpha}$ for $d$=6. (a) large parameter $\tilde{\alpha}$ range and (b) small parameter $\tilde{\alpha}$ range.}\label{pCTd6b}
\end{figure}

The equation of state and the Gibbs free energy are
\begin{eqnarray}
 \tilde{P}&=&\frac{\tilde{r}_{\text{h}}^{4-2d}}{8\pi}-\frac{(d-2)\left(d-4\pi\tilde{T} \tilde{r}_{\text{h}}-3\right)}{16\pi\tilde{r}_{\text{h}}^2}-\frac{\tilde{\alpha}(d-2)\left(d-8\pi \tilde{T}\tilde{r}_{\text{h}}-5\right)}{16\pi\tilde{r}_{\text{h}}^4},\\
 \tilde{G}&=&\frac{(d-2)\omega_{d-2}\tilde{r}_{\text{h}}^{3-d}}{4\pi\left(d^2-4d+3\right)}
 +\frac{\omega_{d-2}\tilde{r}_{\text{h}}^{d-3}\left(d-2\pi\tilde{T}\tilde{r}_{\text{h}}-2\right)}{8\pi  (d-1)}+\frac{\tilde{\alpha}(d-2)\omega_{d-2}\tilde{r}_{\text{h}}^{d-5}\left(d-6\pi\tilde{T}\tilde{r}_{\text{h}}-4\right)}{4\pi (d-4)(d-1)}.
\end{eqnarray}
Solving $(\partial_{\tilde{r}_{\text{h}}}\tilde{P})_{\tilde{T},\tilde{\alpha}}=(\partial_{\tilde{r}_{\text{h}},\tilde{r}_{\text{h}}}\tilde{P})_{\tilde{T},\tilde{\alpha}}=0$, we can obtain the critical point for different $d$. When $d$=6, we plot the critical temperature $\tilde{T}_{\text{c}}$ as a function of $\tilde{\alpha}$ in Fig. \ref{pCTd6b}. Interestingly, for fixed $\tilde{\alpha}$, more than one critical temperatures are observed, which implies that there are several critical points. In a small region $\tilde{\alpha}\in (2.9, 3.6)$ shown in Fig. \ref{CTd6b}, we observe three critical temperatures. So there must be some phase transitions beyond the small-large black holes of the VdW type. Actually, there exists a triple point for the six-dimensional charged AdS black holes, which was first observed in our previous paper \cite{Wei2}. The exact phase diagram for the triple point was also given. Here the triple point is out of our consideration in this paper, so we will leave it for further study.

\begin{figure}
\center{\includegraphics[width=7cm]{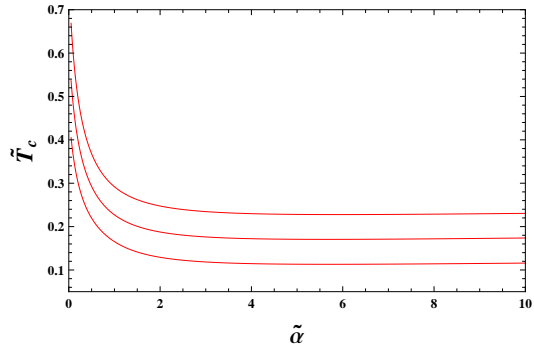}}
\caption{Critical temperature $\tilde{T}_{\text{c}}$ as a function of $\tilde{\alpha}$ for $d$=7, 8, 9 from bottom to top.}\label{CTd789alpha}
\end{figure}

For $d$=7-9, we describe the corresponding critical temperature in Fig. \ref{CTd789alpha}. For each $d$, we find that there is only one critical temperature for fixed $\tilde{\alpha}$. Its value increases with $d$. So in higher dimensions, there exists only the VdW type phase transition. Another interesting phenomenon is that there always exists a lower bound for the critical temperature for each $d$. The locations of the lower bounds ($\tilde{\alpha}$, $\tilde{T}_{\text{c}}$) for $d$=7, 8, 9 are (5.7207, 0.1131), (5.6586, 0.1706), and (5.9334, 0.2277). Obviously, the value of $\tilde{T}_{\text{c}}$ increases with spacetime dimension number $d$. For $\tilde{\alpha}$=0, the critical temperatures keep finite values, which are related to that of the $d$-dimensional charged AdS black holes given in \cite{WeiWeiWei}, where the small-large black hole phase transition was found.

The normalized scalar curvature for any $d$ can be calculated through (\ref{scalarcurv}), which is
\begin{eqnarray}
 R_{\text{N}}&=&\frac{1}{2}-\frac{2\pi^2\tilde{T}^2\tilde{r}_{\text{h}}^{4d+2}\left(6\tilde{\alpha}+\tilde{r}_{\text{h}}^2\right)^2}{\left(\tilde{r}_{\text{h}}^{2 d}\left(\tilde{r}_{\text{h}}^2\left(-d+2\pi \tilde{T}\tilde{r}_{\text{h}}+3\right)-2\tilde{\alpha}\left(d-6\pi \tilde{T} \tilde{r}_{\text{h}}-5\right)\right)+2\tilde{r}_{\text{h}}^8\right)^2}.
\end{eqnarray}
It is easy to check that $R_{\text{N}}$ diverges at the spinodal curve. On the other hand, it vanishes when one of the following conditions is satisfied.
\begin{eqnarray}
 \tilde{r}_{\text{h}}^{2 d-8} \left(\tilde{r}_{\text{h}}^2 \left(-d+4\pi \tilde{T}
   \tilde{r}_{\text{h}}+3\right)-2\tilde{\alpha} \left(d-12\pi \tilde{T}
   \tilde{r}_{\text{h}}-5\right)\right)+2=0,\\
 \tilde{r}_{\text{h}}^{2 d-8} \left(2 \tilde{\alpha} (d-5)+(d-3)
   \tilde{r}_{\text{h}}^2\right)-2=0.
\end{eqnarray}
The first condition depends on $\tilde{T}$, while the second one does not. So in the $\tilde{T}-\tilde{r}_{\text{h}}$ diagram, the first condition gives a curve
\begin{eqnarray}
 \tilde{T}=\frac{2\tilde{\alpha}(d-5)-2\tilde{r}_{\text{h}}^{8-2d}+(d-3)\tilde{r}_{\text{h}}^2}{4\pi \tilde{r}_{\text{h}}\left(6\tilde{\alpha}+\tilde{r}_{\text{h}}^2\right)},
\end{eqnarray}
closely depending on $\tilde{\alpha}$ and $d$. While the second condition shows a specific $\tilde{r}_{\text{h}}$ for each $d$, which is a vertical line in the $\tilde{T}-\tilde{r}_{\text{h}}$ diagram. When fixing $\tilde{\alpha}$=1, we have $\tilde{r}_{\text{h}}$=0.8180, 0.8242, and 0.8347 for $d$=7, 8, and 9, respectively.

For simplicity, we take $\tilde{\alpha}$=1 for following study of the phase diagram and Ruppeiner geometry. The critical points are ($\tilde{P}_{\text{c}}$, $\tilde{T}_{\text{c}}$, $\tilde{r}_{\text{hc}}$)=(0.0184, 0.1105, 1.9138), (0.0448, 0.1653, 1.3999), (0.0954, 0.2271, 1.2171), and (0.1710, 0.2918, 1.1458) for $d$=6-9. The phase diagrams are listed in Fig. \ref{TRhd9d}. We observe that it is similar with that of $d$=5. So for the small black hole with high temperature, $R_{\text{N}}$ is positive, which indicates that the repulsive interaction can exist among the microstructure. This property is different from the neutral case, while similar to that of four-dimensional charged case \cite{WeiWeiWei}.

\begin{figure}
\center{\subfigure[]{\label{TRhd6a}
\includegraphics[width=7cm]{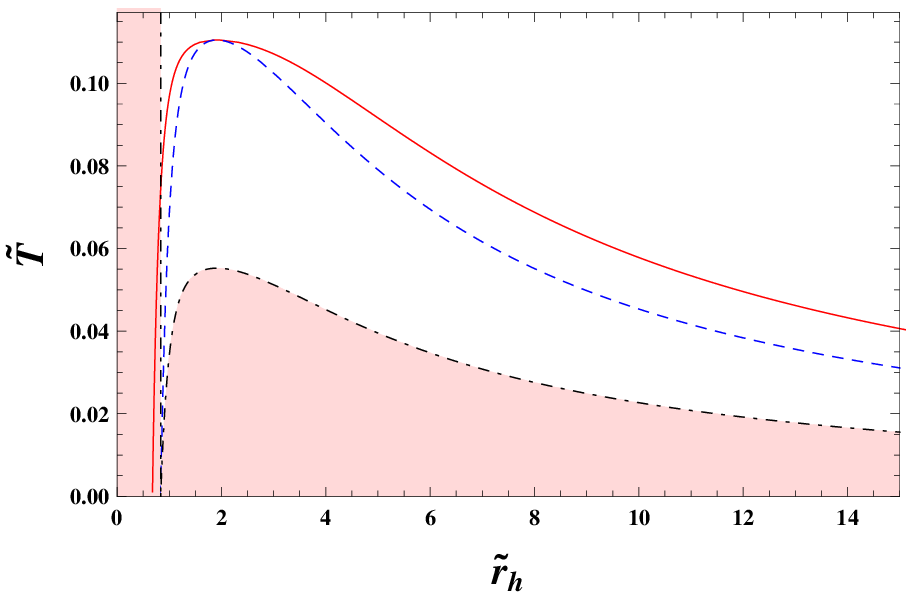}}
\subfigure[]{\label{TRhd7b}
\includegraphics[width=7cm]{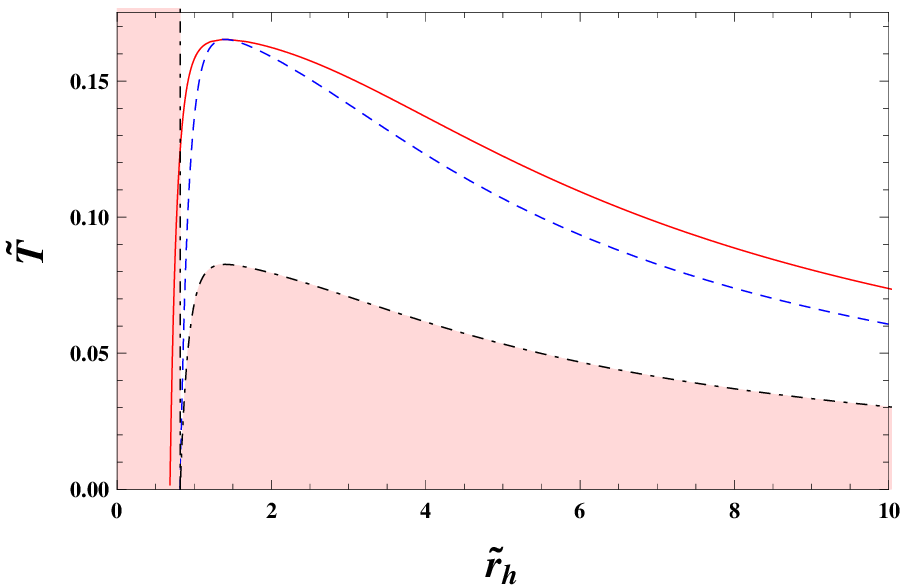}}
\subfigure[]{\label{TRhd8c}
\includegraphics[width=7cm]{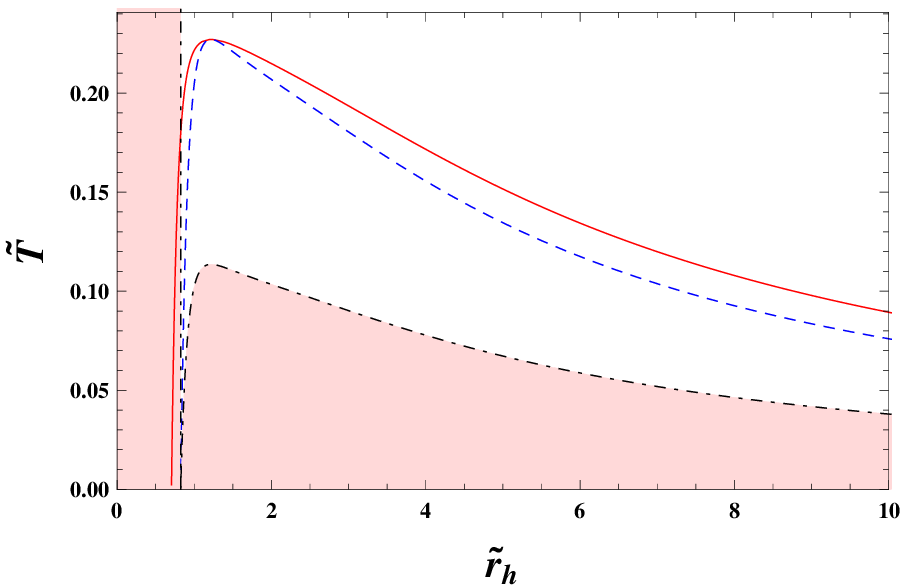}}
\subfigure[]{\label{TRhd9d}
\includegraphics[width=7cm]{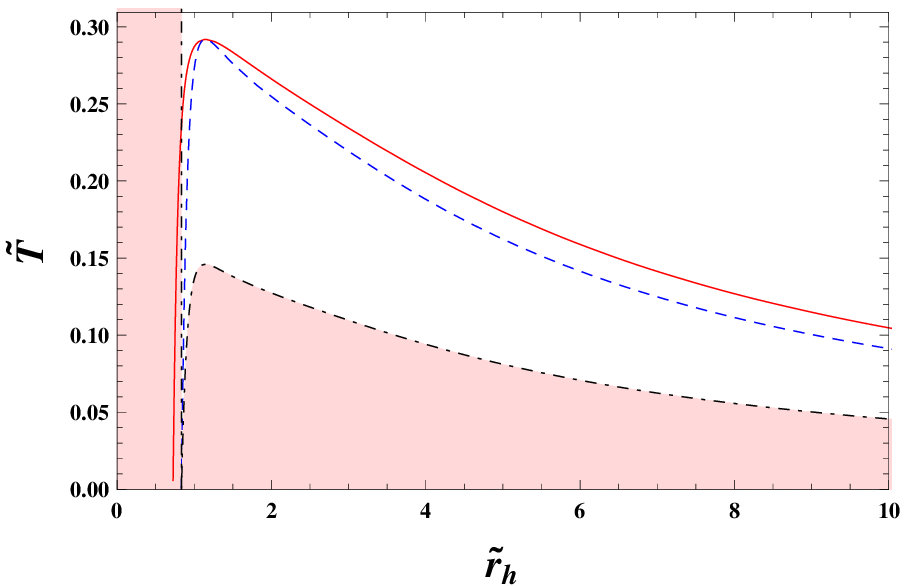}}}
\caption{Phase diagrams for $d$=6, 7, 8, and 9 with $\tilde{\alpha}$=1. The coexistence curves and spinodal curves are marked with the red solid lines and blue dashed lines. The sign-changing curves of $R_{\text{N}}$ is marked with the black dot dashed lines. $R_{\text{N}}$ is positive in the shadow regions, otherwise, it is negative. (a) $d$=6. (b) $d$=7. (c) $d$=8. (d) $d$=9.}\label{TRhd9d}
\end{figure}

Now, we aim to uncover the critical phenomena of the radius of the horizon and the normalized scalar curvatures. We adopt the fitting forms (\ref{fit1}) and (\ref{fit2}). By making use of the numerical calculations, we list the fitting coefficients in Table \ref{highdi}. It is clear that for the coexistence small or large black hole, $b$ and $\beta$ are around 1/2 and -2, respectively. So we can conclude that
\begin{eqnarray}
 \tilde{r}_{\text{h1,2}}\propto(\tilde{T}_{\text{c}}-\tilde{T})^{\frac{1}{2}},\\
 R_{\text{N}}\propto (\tilde{T}_{\text{c}}-\tilde{T})^{-2}
\end{eqnarray}
hold for higher-dimensional black holes. Therefore, based on this result, $\tilde{r}_{\text{h1,2}}$ can act as the order parameter to describe the black hole phase transition. And the normalized scalar curvature also has a critical exponent in higher-dimensional cases.

\begin{table}[h]
\begin{center}
\begin{tabular}{c c c c c c }
  \hline
  \hline
   $d$&$a$ &$b$& $\beta$ & $\gamma$\\
  \hline
 6 & 13.7123/18.2657 &0.4891/0.5110  & -1.9651/-2.0386 &  -6.0817/-6.9303  \\
 7 & 6.5117/10.1555  &0.4829/0.5180  & -1.9683/-2.0385 &  -5.3804/-6.0545  \\
 8 & 3.2645/5.1765   &0.4811/0.5203  & -1.9894/-2.0137 &  -5.0189/-5.0989  \\
 9 & 2.1115/3.5098   &0.4780/0.5241  & -1.9608/-2.0487 &  -4.2470/-4.9299  \\
  \hline\hline
\end{tabular}
\end{center}
\caption{Fitting coefficients for $d$=6-9 with $\tilde{\alpha}$=1. .../... means the values are for coexistence small/large black holes. The fitting forms are given in (\ref{fit1}) and (\ref{fit2}).}\label{highdi}
\end{table}

\section{Conclusions and discussions}
\label{Conclusion}

We have studied the phase diagram and Ruppeiner geometry for the $d$-dimensional charged GB-AdS black holes. The VdW like phase transitions were examined and the normalized scalar curvatures were calculated.

For $d$=5, we obtained an analytical form for the critical point of the phase transition. After reducing these thermodynamic quantities with the black hole charge, we found that the critical point is a curve starting at a finite point and ending at the origin with the increase of $\tilde{\alpha}$ in the $\tilde{P}$-$\tilde{T}$ diagram. Across the curve, the heat capacity exhibits a $\lambda$-like phase transition behavior of $^4$He, which indicates that the phase transition is second order. The phase diagrams were clearly shown in the $\tilde{P}$-$\tilde{T}$ and $\tilde{T}$-$\tilde{r}_{\text{h}}$ diagrams. Moreover, we also observed that the radius of the event horizon can serve as the order parameter to characterize the black hole phase transition.

Then we constructed the Ruppeiner geometry for the five-dimensional charged GB-AdS black holes. The corresponding normalized scalar curvatures are calculated. After excluding the coexistence region, we observed that $R_{\text{N}}$ can be negative or positive. According to the empirical observation of the Ruppeiner geometry, the repulsive interaction is allowed for the microstructure of the small black hole with higher temperature. This result is similar to the four-dimensional charged GB-AdS black holes, while different from the five-dimensional neutral GB-AdS black holes.

For the five-dimensional neutral GB-AdS black holes, one intriguing property is that its microscopic interaction keeps unchanged when the phase transition takes place. However, for the charged cases, when the system goes along the coexistence small and large black holes, we found that their scalar curvatures do not coincide indicating that the microscopic interaction changes during the phase transition, where the black hole microstructure suddenly changes. So the intriguing property does not hold for the charged black holes. We further confirmed that there are the critical phenomena for the normalized scalar curvatures.

Furthermore, we found that for the black holes in more than five dimensions, they share the similar patterns of the phase diagrams and normalized scalar curvatures. For the  small black hole with higher temperature, the repulsive interaction always dominates among their microstructure. Meanwhile, the microscopic interaction changes when the small-large black hole phase transition occurs, quite different from the five-dimensional neutral case.

The study provides us with preliminary properties of the microstructure of the charged black holes in the GB gravity. Next, we will focus on the six-dimensional charged GB-AdS black hole, where a triple point is present. These shall help us to get deep understanding of the GB gravity.

\section*{Acknowledgements}
This work was supported by the National Natural Science Foundation of China (Grants No. 12075103, No. 11675064, No. 11875151, and No. 12047501), the 111 Project (Grant No. B20063), and the Fundamental Research Funds for the Central Universities (No. Lzujbky-2019-ct06).

\end{document}